\iftrue
\documentclass[aps,prl,twocolumn,
			   groupedaddress,superscriptaddress,
			   amsfonts,amssymb,amsmath,
			   citeautoscript,
			   a4paper]{revtex4-1}
\fi

\usepackage[pdftex]{hyperref}
\hypersetup{colorlinks,
			linkcolor={blue!75!black!80!yellow},
			citecolor={blue!75!black!80!yellow},
			urlcolor={blue!75!black!80!yellow},
			pdfstartview=FitH}
\usepackage{graphicx}
\usepackage{mathrsfs}
\usepackage{xspace}
\usepackage{braket}
\usepackage{xr}
\usepackage{gensymb} 
\usepackage{xcite}
\usepackage{xcolor,soul}
\usepackage{stmaryrd} 
\usepackage[UKenglish]{babel}
\usepackage{placeins} 
\usepackage{bbm}
\usepackage{physics}
\usepackage{siunitx}

\DeclareSIUnit[number-unit-product=]\percent{\char`\%} 

\usepackage{txfonts}  
\usepackage{txfontsb} 

\makeatletter
\newcommand*{\addFileDependency}[1]{
  \typeout{(#1)}
  \@addtofilelist{#1}
  \IfFileExists{#1}{}{\typeout{No file #1.}}
}
\makeatother

\newcommand*{\myexternaldocument}[1]{%
    \externaldocument{#1}%
    \addFileDependency{#1.tex}%
    \addFileDependency{#1.aux}%
}

\makeatletter
\renewcommand\@make@capt@title[2]{%
	\@ifx@empty\float@link{\@firstofone}{\expandafter\href\expandafter{\float@link}}%
	\sffamily{\textbf{#1}}\@caption@fignum@sep#2
}%

\makeatother

\newcommand{\appropto}{\mathrel{\vcenter{
			\offinterlineskip\halign{\hfil$##$\cr
				\propto\cr\noalign{\kern2pt}\sim\cr\noalign{\kern-2pt}}}}}

\newcommand{\EXP}{\text{e}}

\newcommand{\I}{\mathrm{i}}

\newcommand{\omegap}{\omega_{\rm{p}}}

\newcommand{\gquplus}{g_\mathrm{Q}^{(+)}}
\newcommand{\gquminus}{g_\mathrm{Q}^{(-)}}
\newcommand{\gplus}{g^{(+)}}
\newcommand{\gminus}{g^{(-)}}
\newcommand{\gqu}{g_\mathrm{Q}}
\newcommand{\kplus}{k^{(+)}}
\newcommand{\kminus}{k^{(-)}}
\newcommand{\omegaplus}{\omega^{(+)}}
\newcommand{\omegaminus}{\omega^{(-)}}

\newcommand{\barePhOp}{\hat{a}}
\newcommand{\SqPhOp}{\hat{a}_\mathrm{s}}
\newcommand{\bplus}{{\hat{b}^{(+)}}}
\newcommand{\bminus}{\hat{b}^{(-)}}
\newcommand{\Ssq}{\hat{S}_{\mathrm{s}}}
\newcommand{\Scattbogoplus}{\hat{S}^{(+)}}
\newcommand{\Scattbogominus}{\hat{S}^{(-)}}

\newcommand{\insetZTX}[1]{\ifbool{togglechanges}
    {#1}  
    {\textcolor{cyan!60!green}{#1}}}

\usepackage{textcomp} 
\usepackage{xifthen}
\usepackage{etoolbox}
\newboolean{togglecomments}
\newboolean{togglechanges}

\setboolean{togglecomments}{false}
\setboolean{togglechanges}{true}

\newcommand{\comment}[2]{%
    \ifbool{togglecomments}%
    {\textcolor{blue!70!black}{\small\textsf{%
    \textsuperscript{\textsc{\textsf{\MakeLowercase{#1}}}}%
    [#2]}}} 
    {}}     
\newcommand{\swap}[2]{\ifbool{togglechanges}
    {#2}  
    {\textcolor{red!70!black}{[#1]}\textrightarrow{}\textcolor{green!50!black}{[#2]}}}
\newcommand{\remove}[1]{\ifbool{togglechanges}
    {}    
    {\textcolor{red!70!black}{#1}}}
\newcommand{\inset}[1]{\ifbool{togglechanges}
    {#1}  
    {\textcolor{green!50!black}{#1}}}

\newcommand{\optional}[1]{\ifbool{togglechanges}
    {}    
    {\textcolor{yellow!50!orange!80!gray}{#1}}}

\newcommand{\citeremind}[1]{%
    [\textcolor{blue!75!black!80!yellow}{
        $\blacksquare$%
	    \ifthenelse{\isempty{#1}}
	        {}
	        {\textsuperscript{\tiny\textsf{#1}}}%
	}]\xspace}

\newcommand{\ie}{i.e.\@\xspace}  

\newcommand{\eg}{e.g.\@\xspace}

\newcommand{\hkuaffil}{\footnotesize Department of Physics and HK Institute of Quantum Science and Technology,
The University of Hong Kong, Pokfulam, Hong Kong, China}
\newcommand{\hkustate}{\footnotesize State Key Laboratory of Optical Quantum Materials, The University of Hong Kong, Pokfulam, Hong Kong, China}

\myexternaldocument{SM}

\begin{document}

\title{
Optical parametric free-electron--photon quantum interaction 
}

\author{Zetao~Xie}
\affiliation{\hkuaffil}
\affiliation{\hkustate}
\author{Zehai~Pang}
\affiliation{\hkuaffil}
\affiliation{\hkustate}
\author{Yi~Yang}
\email{yiyg@hku.hk}
\affiliation{\hkuaffil}
\affiliation{\hkustate}

\begin{abstract}
Optical parametric processes underpin quantum photonics, while free-electron--photon interactions offer agile pathways to generate nontrivial quantum photonic states. 
These threads have so far largely progressed independently, whereas placing free electrons in a driven nonlinear system can potentially activate coherent parametric interaction channels for joint state engineering of both types of particles.
Here we unify these paradigms by developing a general theoretical framework for parametric free-electron--photon interactions in a nonlinear optical system driven by degenerate parametric down-conversion.
Unlike free electrons in a linear bath, here they can couple to Bogoliubov quasiparticles through two detuned phase-matching channels, where the parametric process and free-electron interactions can quantum amplify each other.
Seeding the interaction with squeezed vacuum yields gain-only or loss-only electron energy spectra, and enables electron-heralded squeezed Fock states;
with bare vacuum, postselecting electron energy sidebands generates high-fidelity Schrödinger cat states.
Our results show how optical parametric interactions can quantum shape free electrons and photons, potentially enabling a quantum parametric dielectric laser accelerator that mitigates the need for temporal phase synchronization, thereby allowing acceleration probabilities to approach unity even for phase-random electrons.

\end{abstract}

\maketitle

Optical parametric processes have long played a central role in nonlinear optics and quantum optics. 
Early table-top optics based on the second-order $\chi^{(2)}$ and
third-order $\chi^{(3)}$ parametric interactions laid the foundation for optical parametric oscillation~\cite{giordmaine1965tunable,pepper1978observation} and amplification~\cite{wang1965measurement,carman1966observation}, and pioneered the generation of squeezed light~\cite{slusher1985observation, wu1986generation,shelby1986broad} and entangled photon pairs~\cite{burnham1970observation, kwiat1995new}.
With the rise of integrated photonics, parametric interactions now power scalable quantum technologies, including integrated quantum sources, on-chip quantum communication, quantum information processing, and the generation of complex non-Gaussian states~\cite{wang2020integrated, zhu2021integrated,dutt2024nonlinear}.
Interfacing these boson-mediated interactions with matter in cavities allows the engineering and enhancement of light-matter interaction~\cite{lu2015squeezed,zeytinouglu2017engineering,qin2018exponentially,sanchez2021squeezed,le2025cavity}, creating opportunities for entering the ultrastrong and even deep-strong coupling regimes~\cite{forn2019ultrastrong, qin2024quantum}.

These regimes are common essential goals in studying the interplay between free electrons and photons, an important type of light--matter interaction leading to fundamental discoveries~\cite{cerenkov1934visible, friedman1988spontaneous,garcia2010optical} and technological development~\cite{liu2017integrated, polman2019electron,shiloh2022miniature,yang2023photonic,roques2022framework,roques2023free,shi2025quantum}.
In recent years, the invention of photon-induced near-field electron microscopy (PINEM)~\cite{barwick2009photon} and its quantum description~\cite{garcia2010multiphoton, park2010photon} have stimulated the emerging field of free-electron quantum optics~\cite{talebi2018electron,rivera2020light,garcia2021optical,ruimy2025free,garcia2025roadmap}. 
The theoretical description of PINEM was subsequently generalized into a fully second-quantized form---quantum PINEM (QPINEM)~\cite{kfir2019entanglements,di2019probing,ben2021shaping}.
This generalization describes how the interaction generates entanglement between electrons and photons~\cite{kfir2019entanglements,di2019probing,ben2021shaping,henke2025observation, preimesberger2025experimental}, and among electrons~\cite{kfir2019entanglements, kumar2024strongly} and photons~\cite{baranes2022free} themselves.
PINEM interactions can enable applications like ultrafast free-electron probe~\cite{piazza2015simultaneous,wang2020coherent,kfir2020controlling,henke2021integrated,kurman2021spatiotemporal,yang2024free}, quantum shaping of free electrons~\cite{feist2015quantum,morimoto2018diffraction,pan2019anomalous, vanacore2019ultrafast,dahan2021imprinting,nabben2023attosecond,gaida2024attosecond,fang2024structured}, and quantum light generation~\cite{bendana2011single,feist2022cavity,dahan2023creation,arend2024electrons}.
These prospects of free-electron quantum optics complement pure optical parametric schemes with broad spectral agility, tunable electron-photon phase matching, and energy-resolved postselection.

It is thus enticing to consider various types of nonlinearities in the QPINEM process.
These nonlinearities fall into three classes according to their physical origin.
The first class stems from light--matter coupling, including higher-order contributions in the minimally coupled Hamiltonian such as the Kapitza-Dirac effect~\cite{batelaan2007colloquium} and other ponderomotive interactions~\cite{kozak2018inelastic,kozak2018ponderomotive,garcia2021opticalprl,di2022optical,tsarev2023nonlinear}.
The second class arises from the non-negligible nonlinear dispersion of low-energy free electrons, where quantum recoil occurs after each single photon emission or absorption event~\cite{talebi2020strong,garcia2022complete,karnieli2023jaynes,eldar2024self,karnieli2024strong,karnieli2024universal,sirotin2024quantum,synanidis2024quantum}.
The third class, most relevant to this work, relies on the nonlinear optical response of the photonic subsystem.
Intrinsic $\chi^{(2)}$~\cite{konecna2020nanoscale,prelat2025wave} and  $\chi^{(3)}$~\cite{cox2020nonlinear, yang2024free,karnieli2024universal} susceptibilities mediate coupling between free electrons and nonlinear optical fields, with the nonlinear response tailoring electron energy and photon emission spectra.
Considering the prominence of both pure optical parametric processes and free-electron interactions in quantum optics, an intriguing open question is whether their synergy can be harnessed to develop quantum light sources and free-electron quantum applications.

\begin{figure*}[htbp]
	\centering
	\includegraphics[width=1\linewidth]{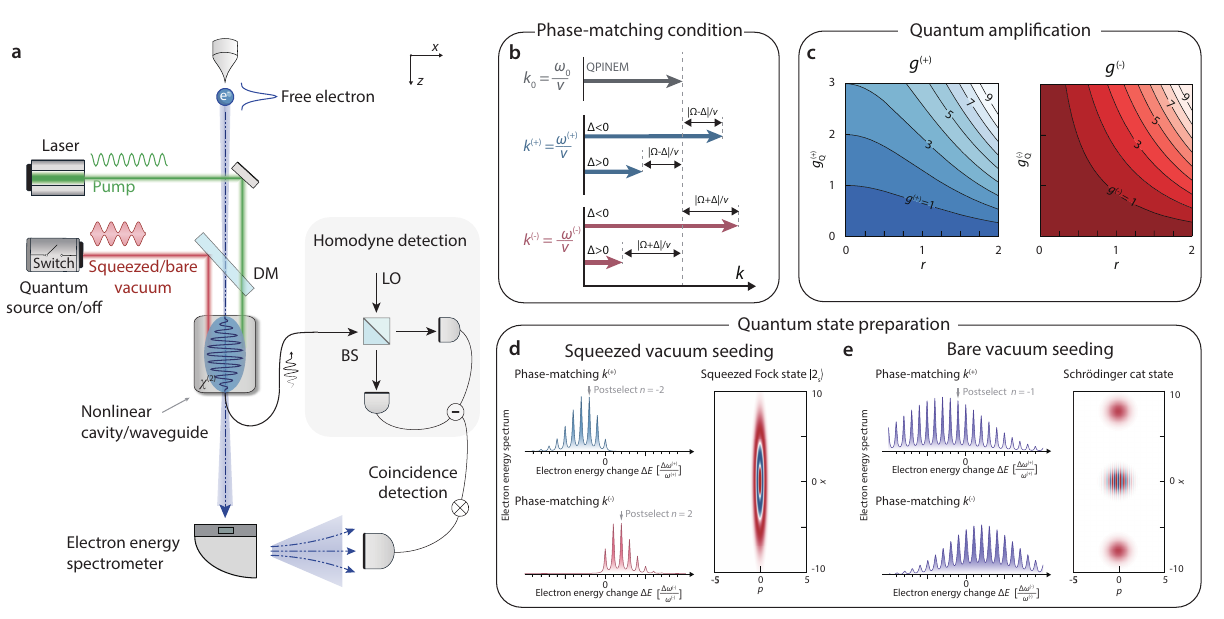}
	 \caption{%
	 	\textbf{Theoretical framework.} 
	 \textbf{a.} Schematic of free-electron--photon interaction driven by DPDC. 
     A nonlinear cavity or waveguide with mode frequency $\omega_0$ is parametrically driven by an undepleted pump at frequency $\omegap$ through nonlinear $\chi^{(2)}$-mediated down-conversion. 
     A quantum source seeds the cavity with an initial photonic state of either a squeezed vacuum (source on) or the bare vacuum (source off).
     A free electron interacts with the nonlinear system and is then measured with an energy spectrometer.
     The output photon state is characterized via homodyne detection in coincidence with the postselected electron energy sideband.
     DM--Dichroic mirror. BS--Beam splitter. LO--Local oscillator.
     %
     \textbf{b.} Parametric phase-matching condition. 
     Existing QPINEM requires the phase velocity of the mode at $\omega_0$ to match the free-electron velocity $v_z$, \ie, the phase-matching momentum is $k_0 = \omega_0/ v_z $ (gray arrow).
     With DPDC driving, the phase-matching momentum is shifted to $\kplus$ (blue arrows) with detuned momentum $\abs{\Omega - \Delta}/v_z$, or to $\kminus$ (red arrows) with detuned momentum $\abs{\Omega + \Delta}/v_z$, where $\Omega$ is the squeezed mode frequency.
     %
     \textbf{c.} $g^{(\pm)}$ as a function of $\gqu^{(\pm)}$ and squeezing parameter $r$. 
    The interaction strength between the free electron and the squeezed mode is given by $\gplus =  \gquplus \, \cosh{r}$ or $\gminus =\gquminus \, \sinh{r}  $ for the phase-matching condition $\kplus$ or $\kminus$, respectively, both of which exhibit an exponential enhancement for $r>1$.
     %
     \textbf{d.} Squeezed Fock state preparation. 
     When the interaction is seeded by squeezed vacuum, postselection on free-electron sidebands [either gain-only sidebands for $\kplus$ (top left) or loss-only sidebands for $\kminus$ (bottom left)] heralds the generation of squeezed Fock states (right).
     \textbf{e.} Cat state preparation. 
     When the interaction is seeded by bare vacuum, postselection on free-electron sidebands [gain or loss sidebands for either $\kplus$ (top left) or $\kminus$ (bottom left) phase matching] heralds the generation of Schrödinger cat states (right; an odd cat state with magnitude $\alpha = 5.431$).
     Even and odd electron energy sidebands correspond to even and odd cat states, respectively.
     In (\textbf{d}) and (\textbf{e}), ${\gqu^{(\pm)}}=1$, the squeezing parameter ${r}=1.15$ ($\approx\SI{10}{dB}$), detuning $\Delta=0.2\omega_0$ is chosen for better visualizing the electron energy shifts, and energy spectra are normalized to $\hbar\omega^{(\pm)}$ with a Lorentzian broadening of $0.3\omega^{(\pm)}$.
             }      
	\label{fig_main:framework}
\end{figure*}

Here, we develop a theoretical framework for the quantum interaction between free electrons and photons mediated by a degenerate parametric down-conversion (DPDC) process [Fig.~\ref{fig_main:framework}(a)].
We predict that the parametric driving splits the original electron-photon phase-matching condition into two detuned ones [Fig.~\ref{fig_main:framework}(b)], and identify a squeezing-enabled quantum amplification of the interaction strength [Fig.~\ref{fig_main:framework}(c)].
Seeded by a squeezed vacuum, the interaction produces distinctive single-sided loss/gain electron energy spectra, and generates a one-to-one entanglement between electron sidebands and squeezed Fock states [left panel in Fig.~\ref{fig_main:framework}(d)].
Meanwhile, seeded by a bare vacuum, the interaction enables the high-fidelity generation of Schrödinger cat states with cat amplitudes highly tunable via electron-photon and photon-photon coupling [right panel in Fig.~\ref{fig_main:framework}(d)] .
Finally, an enticing application of the framework is a new type of quantum parametric dielectric laser accelerators that mitigates the stringent temporal phase synchronization requirement, enabling near-unity acceleration probability.

\subsection{Theoretical framework}

The Hamiltonian describing the system in Fig.~\ref{fig_main:framework}(a) is generally governed by
\begin{align}
    \hat{H} = \hat{H}_0 +\hat{H}_\text{el-ph} + \hat{H}_\text{DPDC},
    \label{eq:total-general-H}
\end{align}
where the free Hamiltonian $\hat{H}_0 = \hat{H}_\text{el} +  \hat{H}_\text{ph} $ consists of the free-electron and photonic parts, $\hat{H}_\text{el-ph}$ describes their interaction, and $\hat{H}_\text{DPDC}$ corresponds to the degenerate parametric down-conversion (DPDC) process. 
The specific Hamiltonian components are (see SM Sec. S1)
\begin{equation}
\begin{aligned}
&  \hat{H}_\text{el}  = -\I \hbar \mathbf{v} \cdot \nabla ,\\
&\hat{H}_\text{ph} = \hbar \omega_0 \barePhOp^\dagger \barePhOp,\\
& \hat{H}_\text{el-ph} = {e  \mathbf{v}} \cdot \left[   {\mathbf{A}}(\mathbf{r})\barePhOp + {\mathbf{A}}^* (\mathbf{r})\barePhOp^\dagger  \right],  \\
& \hat{H}_\text{DPDC} =  - \hbar \eta \left(\EXP^{- \I \omegap t} \barePhOp^{\dagger 2} +  \EXP^{ \I \omegap t} \barePhOp^{ 2} \right).
    \label{eq:total-H}
\end{aligned}
\end{equation}
Here $\barePhOp$ and $\barePhOp^\dagger$ are the noncommuting annihilation and creation operators of the photonic structure mode with resonance frequency $\omega_0$, $\mathbf{v}$ is the electron velocity, and the vector potential is given by ${\mathbf{A}}(\mathbf{r}) = {\mathbf{E}}(\mathbf{r})/\I \omega_0  $ with ${\mathbf{E}}(\mathbf{r})$ denoting the electric field.
The parametric drive strength is defined as $\eta =\sqrt{\overline{N}_\text{p}} \eta_0 $, where $\overline{N}_\text{p}$ is the mean photon number of the pump with frequency $\omegap$, $\eta_0$ is the down-conversion rate describing the effective intermode coupling, which is proportional to the second-order susceptibility $\chi^{(2)}$~\cite{boyd2008nonlinear}.

The Hamiltonian is derived under a few approximations (see SM Sec. S1).
On the electron side, we take the paraxiality and non-recoil approximation for free electrons, and assume that the magnetic vector potential is weak compared to the electron momentum, thereby neglecting the ponderomotive potential in the free-electron--light interaction~\cite{kfir2019entanglements,di2019probing,ben2021shaping}.
On the photon side, we apply the parametric approximation by replacing the pump operator with a classical field for the DPDC process~\cite{scully1997quantum, navarrete2022introduction}.
In addition, we assume that free electrons are phase-matched with the down-conversion photons at $\omega_0$ but not the pump photons at $\omegap$ such that the electron-photon interaction takes place under the single-mode condition.

To describe the evolution of the quantum system governed by the Hamiltonian Eq.~\eqref{eq:total-general-H}, we analytically derive the scattering operator connecting the initial and final states. 
We first transform Eq.~\eqref{eq:total-H} to a rotating frame by defining a unitary operator $\hat{U}_\mathrm{c}(t) \equiv \hat{U}_\mathrm{ph}(t)\hat{U}_\mathrm{el}(t) = \mathrm{exp}( -\I t \frac{\omegap}{2} \barePhOp^\dagger \barePhOp ) \mathrm{exp} ({-v_z t \partial_z}) $, yielding (see SM Sec. S2.A)
\begin{equation}
\begin{aligned}
\hat{H}_\mathrm{I} &= \hbar\Delta  \barePhOp^\dagger \barePhOp - \hbar \eta (\barePhOp^{\dagger 2} + \barePhOp^2) \\
& \quad + \frac{ e  v_z}{\I \omega_0} \left[   {\mathrm{E}}_z(z +v_z t) \barePhOp  \EXP^{-\I\omegap t/2} {-}   {\mathrm{E}}_z^* (z +v_z t) \barePhOp^\dagger \EXP^{\I\omegap t/2}  \right].
\label{eq:H_I_DPDC-PINEM}
\end{aligned}
\end{equation}
Here, $\Delta = \omega_0 - \omegap/2$ denotes the slight detuning between the cavity mode and the degenerate down-converted photon pairs.

We diagonalize the purely photonic part via the Bogoliubov transformation $\SqPhOp = \Ssq^\dagger(r) \barePhOp \Ssq(r)$ with a squeezing unitary operator $\Ssq = \exp[ r(\barePhOp^2 - \barePhOp^{\dagger 2} )/2 ] $ and squeezing parameter $r$~\cite{scully1997quantum, navarrete2022introduction}.
Rewriting the Hamiltonian in terms of $\SqPhOp$ gives the expression in the squeezed basis $\ket{n_{\mathrm{s}}}$ (see SM Sec. S2.B):
\begin{equation}
\begin{aligned}
\hat{H}_\mathrm{I} & = \hbar\Omega \, \SqPhOp^\dagger \, \SqPhOp + \hbar[\Omega - \Delta]/2\\
& \quad +  \frac{ e  v_z}{\I \omega_0}  \Big[ {\mathrm{E}}_z(z +v_z t) \, \left(\SqPhOp \cosh{r} + \SqPhOp^\dagger \sinh{r} \right) \EXP^{-\I\omegap t/2}\\
& \quad  {-}  {\mathrm{E}}_z^* (z +v_z t) \, \left( \SqPhOp^\dagger \cosh{r} + \SqPhOp \sinh{r} \right) \EXP^{\I\omegap t/2} \Big] , 
\label{eq:H_Bogo_DPDC-PINEM}
\end{aligned}
\end{equation}
where $r = \frac{1}{4} \ln \left[{(\Delta + 2\eta )/(\Delta -2\eta)} \right] $ is chosen to eliminate the quadratic terms proportional to $\SqPhOp^2$ and $\SqPhOp^{\dagger 2}$; $\Omega = \text{sign}(\Delta) \sqrt{\Delta^2 - (2\eta)^2}$ ( $ \abs{\Delta} > 2\eta$) is the squeezed mode frequency.

Transforming Eq.~\eqref{eq:H_Bogo_DPDC-PINEM} into the interaction picture allows us to obtain a closed-form expression for the scattering operator, which can be written as (see SM Sec. S2.C-Sec. S2.E)
\begin{equation}
\begin{aligned}
\hat{S}
& =  \mathrm{exp} \bigg\{ \cosh{r} \left[  {\gquplus}^* \EXP^{- \I \frac{{\omegap}/{2} + \Omega  }{v_z}  {z}  }   \SqPhOp^\dagger -   \, {\gquplus} \EXP^{ \I \frac{{\omegap}/{2} + \Omega  }{v_z} {z}  } \SqPhOp \right] \\
& \quad - \sinh{r}  \left[    \gquminus \EXP^{ \I \frac{{\omegap}/{2} - \Omega  }{v_z} {z} } \SqPhOp^\dagger - {\gquminus}^* \EXP^{ -\I \frac{{\omegap}/{2} - \Omega  }{v_z} {z}  } \SqPhOp \right] \bigg\},
 \label{eq:S_DPDC-PINEM}
\end{aligned}
\end{equation}
$\hat{S}$ contains two types of dimensionless quantum interaction strengths between free electrons and Bogoliubov quasiparticles:
\begin{subequations}\label{eq:g+&-}
\begin{align}
& \gplus = \gquplus \,\cosh{r} ,  
\label{eq:gplus} \\ 
& \gminus =\gquminus \, \sinh{r} ,
\label{eq:gminus}
\end{align}
\end{subequations}
where
\begin{subequations}\label{eq:gqu+&-}
\begin{align}
& \gquplus = \frac{e }{ \hbar \omega_0}     \int_{-\infty}^{\infty} {\mathrm{d} z^\prime} {\mathrm{E}}_z(z^\prime ) \EXP^{ -\I \frac{ {\omegap}/{2} + \Omega }{v_z}  {z^\prime}  },  
\label{eq:gquplus} \\ 
& \gquminus =\frac{e }{ \hbar \omega_0} \int_{-\infty}^{\infty} \mathrm{d} z^\prime   {\mathrm{E}}_z(z^\prime ) \EXP^{ - \I \frac{{\omegap}/{2} - \Omega  }{v_z}  {z^\prime} }.
\label{eq:gquminus}
\end{align}
\end{subequations}
$\gplus$ and $\gminus$ require, respectively, the phase-matching conditions 
\begin{subequations}\label{eq:k+&-}
\begin{align}
&\kplus=  \left(\frac{\omegap}{2} + \Omega\right) {/v_z} ,
\label{eq:kplus} \\ 
&\kminus =  \left(\frac{\omegap}{2} - \Omega\right) {/v_z},
\label{eq:kminus}
\end{align}
\end{subequations}
where $k^{(\pm)}$ are implicitly contained in $ {\mathrm{E}}_z(z^\prime ) $ in Eq.~\eqref{eq:gqu+&-} and we define $\omegaplus \equiv \frac{\omegap}{2} + \Omega $ and $\omegaminus \equiv \frac{\omegap}{2} - \Omega $ for convenient notation henceforth.
The operators $\EXP^{ \I \frac{\omegaplus }{v_z}}$ and $\EXP^{ \I \frac{\omegaminus }{v_z}}$ in Eq.~\eqref{eq:S_DPDC-PINEM} can be further regarded as electron energy ladder operators~\cite{feist2015quantum,kfir2019entanglements,ben2021shaping} with different energy spacing $\hbar \omegaplus$ and $\hbar \omegaminus$, respectively, which we denote as $\bplus{^\dagger}$ and  $\bminus{^\dagger}$.

The interaction strengths Eq.~\eqref{eq:g+&-} consist of two distinct contributions: $\gquplus$ and $\gquminus$, analogous to the spontaneous coupling strength with vacuum $\gqu$ in existing QPINEM theory~\cite{kfir2019entanglements,di2019probing,ben2021shaping,xie2025maximal,zhao2025upper}, originating from the interaction between the free electron and the vacuum photon; $\cosh{r}$ and $\sinh{r}$, which stem from the two-photon process in DPDC.
In the absence of nonlinearity, \ie, $r = 0$, the scattering matrix Eq.~\eqref{eq:S_DPDC-PINEM}, the coupling strength Eq.~\eqref{eq:g+&-}, and the phase-matching conditions Eq.~\eqref{eq:k+&-}, all reduce to their conventional counterparts in the existing QPINEM theory (see Sec. S2.E).

A compelling consequence of the parametric process is the modification of the phase-matching condition [Fig.~\ref{fig_main:framework}(b)] in the free-electron--photon interaction.
In the existing QPINEM, the phase-matching condition requires that the phase velocity of the optical mode at $\omega_0$ matches the free-electron velocity $v_z$, yielding the momentum $k_0 = \omega_0/v_z$ [grey arrow in Fig.~\ref{fig_main:framework}(b)].
Under the nonlinear DPDC driving, however, the phase-matching momentum for the mode $\omega_0$ is fundamentally altered to Eqs.~\eqref{eq:k+&-}
$k^{(\pm)} = \omega^{(\pm)}/v_z$ [depicted by the blue and red arrows in Fig.~\ref{fig_main:framework}(b), respectively].
This phase-matching shift directly manifests in the electron energy spectrum, with the sideband spacing given by $\hbar\omegaplus$ or $\hbar\omegaminus$, which is compressed ($\Delta>0$) or expanded ($\Delta<0$) compared to the spacing $\hbar\omega_0$ in the existing QPINEM (elaborated in the following sections).
Note that here the phase matching undergoes a universal change regardless of the number of particles in the cavity.

Because the phase-matching conditions for $\kplus$ and $\kminus$ cannot be simultaneously satisfied under the single-mode condition,
the full scattering matrix Eq.~\eqref{eq:S_DPDC-PINEM} splits into two separate possibilities, each corresponding to the phase-matching conditions $\kplus$ and $\kminus$:
\begin{subequations}\label{eq:S_bogo_two}
\begin{align}
&\Scattbogoplus
 =  \mathrm{exp}  \left[  {\gplus}^{*} \bplus   {\SqPhOp}^{\dagger} -   \, {\gplus} \bplus{^\dagger} \SqPhOp \right], 
\label{eq:S_bogo_plus} \\
&\Scattbogominus 
= \mathrm{exp}  \left[  {\gminus}^{*} \bminus \SqPhOp  - \gminus \bminus{^\dagger}   {\SqPhOp}^{\dagger}  \right].
\label{eq:S_bogo_minus}
\end{align}
\end{subequations}
Here, Eq.~\eqref{eq:S_bogo_plus} resembles the existing QPINEM processes, where the electron energy gain accompanies the annihilation of a Bogoliubov quasiparticle.
In contrast, Eq.~\eqref{eq:S_bogo_minus} describes the simultaneous electron energy gain and the creation of Bogoliubov quasiparticles, which, as we will show below, have its unique application in dielectric laser accelerators (DLAs; see Fig.~\ref{fig_main:accelerator}).

Aside from the phase-matching modification, another crucial aspect is the quantum amplification of the free-electron--photon interaction.
Compared to the interaction strength $\gqu$ in the existing QPINEM, $\gquplus$ and $\gquminus$, can be boosted by $\cosh{r}$ and $\sinh{r}$ for the interaction between free electron and Bogoliubov quasiparticle, respectively, providing exponential enhancement $\sim \exp(r)/2$ for $r>1$ [Fig.~\ref{fig_main:framework}(c)].
This quantum amplification stems from the parametric process that converts $\sim \exp(2r)/4$ photons into the Bogoliubov quasiparticles.

Our framework enables the generation of non-Gaussian photonic states, including squeezed Fock states and Schrödinger cat states.
With a squeezed vacuum input [Fig.~\ref{fig_main:framework}(d)],   
postselecting the electron energy loss sideband for $\kplus$ phase matching, or the electron gain sideband for $\kminus$, heralds the creation of a squeezed Fock state whose Fock number is in one-to-one correspondence with the electron sideband index (see ``Squeezed vacuum seeding for squeezed Fock states" section).
If we turn off the input squeezed vacuum [Fig.~\ref{fig_main:framework}(e)] in both phase-matching $k^{(\pm)}$, postselecting odd (even) electron sidebands yields odd (even) cat states, where the cat amplitude $\alpha$ increases with the squeezing level $r$ and $\gqu^{(\pm)}$
(see ``Bare vacuum seeding for Schrödinger cat states" section).
Taken together, our framework gives rise to a rich variety of predictions, ranging from parametric electron-photon phase-matching, quantum-amplified interaction strength, and squeezed Fock and cat states generation.

\subsection{Squeezed vacuum seeding for squeezed Fock states}

\begin{figure}[htbp]
	\centering
	\includegraphics[width=1\linewidth]{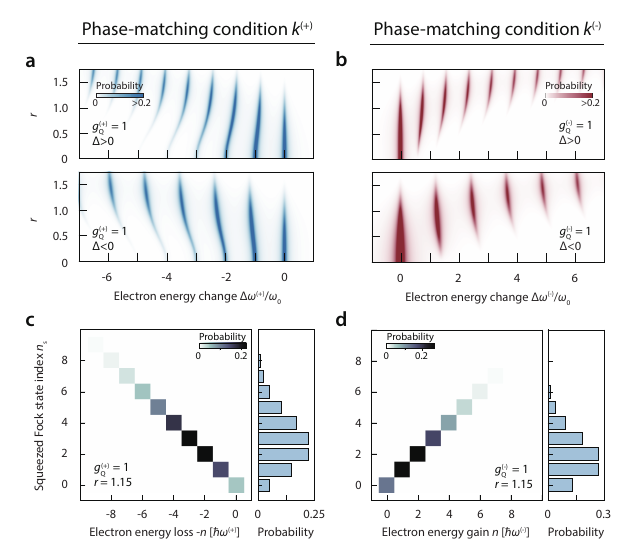}
	 \caption{%
	 	\textbf{Free-electron parametric interaction with squeezed vacuum and its generation of squeezed Fock states.} 
        %
	  \textbf{a-b.} Electron energy spectra as a function of squeezing parameter ${r}$ for phase-matching conditions $\kplus$  (\textbf{a}) and $\kminus$ (\textbf{b}).
      For $\kplus$, only electron energy loss occurs, whereas for $\kminus$, only electron energy gain is observed.
      %
      \textbf{c-d.} Joint probability distribution between the squeezed Fock state and electron energy loss $-n\hbar\omegaplus$ (\textbf{c}) or energy gain $n\hbar\omegaminus$ (\textbf{d}), with nonzero probability only along the anti-diagonal (\textbf{c}) and diagonal (\textbf{d}).
      The right insets show the corresponding Poisson probability distributions of the squeezed Fock states.
      %
      The squeezing parameter is $r=1.15$; plots  (\textbf{a}) and  (\textbf{c}) use $\gquplus=1$ [$\kplus$], and plots  (\textbf{b}) and  (\textbf{d}) use $\gquminus=1$ [$\kminus$]; the detuning is fixed at $\abs{\Delta} = 0.2\omega_0$ and the Lorentzian peak broadening is $0.1\omega^{(\pm)}$. 
        }
	\label{fig_main:quantumsouruce_on}
\end{figure}

We first consider the photonic part of the initial state being the squeezed vacuum, as provided by the quantum source that is switched on [see Fig.~\ref{fig_main:framework}(a)].
The quantum source prepares a squeezed vacuum state $\ket{0_\mathrm{s}}$, allowing the photonic state to be described in the squeezed basis.
The free electron state $\ket{E_n}$ has energy $E_0 + n\hbar\omega$, where $n$ labels the energy ladder and represents energy gain ($n>0$) or loss ($n<0$) relative to the baseline energy $E_0$.
The initial state of the system is therefore given by the joint state of the electron at the baseline energy and the squeezed vacuum state, $\ket{\Psi_{\mathrm{i}}} =\ket{E_0}\otimes \ket{0_\mathrm{s}} \equiv \ket{E_0, 0_\mathrm{s}}$.
The evolution under the interaction is governed by the scattering matrix Eq.~\eqref{eq:S_bogo_two}, yielding the explicit final state for the phase-matching conditions $\kplus$ and $\kminus$, respectively (see Sec. S3.A):
\begin{subequations}\label{eq:Psi_f_bogo_two}
\begin{align}
&\ket{\Psi_{\mathrm{f}}}^{(+)}
= \sum_{n=0}^{\infty} \EXP^{- \frac{ \lvert{\gplus}\rvert^2   }{2} }\frac{ \left[  {\gplus}^{*} \right]^n }{\sqrt{n!}} \ket{E_{-n}, n_\mathrm{s}},
\label{eq:Psi_f_bogo_plus} \\
&\ket{\Psi_{\mathrm{f}}}^{(-)}
= \sum_{n=0}^{\infty} \EXP^{ - \frac{ \lvert{\gminus}\rvert^2    }  {2} } \frac{ \left[  -  \gminus \right]^n }{\sqrt{n!}} \ket{E_{n},  n_\mathrm{s}}.
\label{eq:Psi_f_bogo_minus}
\end{align}
\end{subequations}

The joint state Eq.~\eqref{eq:Psi_f_bogo_plus} describes a correlated superposition in which the number $n_\mathrm{s}$ of the squeezed Fock state exactly equals $n$ quanta of electron energy loss $\hbar\omegaplus$ ($n_\mathrm{s} = n$), with probabilities following a Poisson distribution of expectation $ \lvert{\gplus}\rvert^2$.
By contrast, Eq.~\eqref{eq:Psi_f_bogo_minus} implies the number $n_\mathrm{s}$ of the squeezed Fock state coincides with $n$ quanta of electron energy gain $\hbar\omegaminus$ ($n_\mathrm{s} = n$), with probabilities also Poissonian but of a different expectation $ \lvert{\gminus}\rvert^2$.
The one-to-one correspondence between the electron energy change $\pm n$ and the squeezed Fock state number $n_\mathrm{s}$ signifies the entanglement between the electron and the squeezed Fock state.

\begin{figure*}[htbp]
	\centering
	\includegraphics[width=1\linewidth]{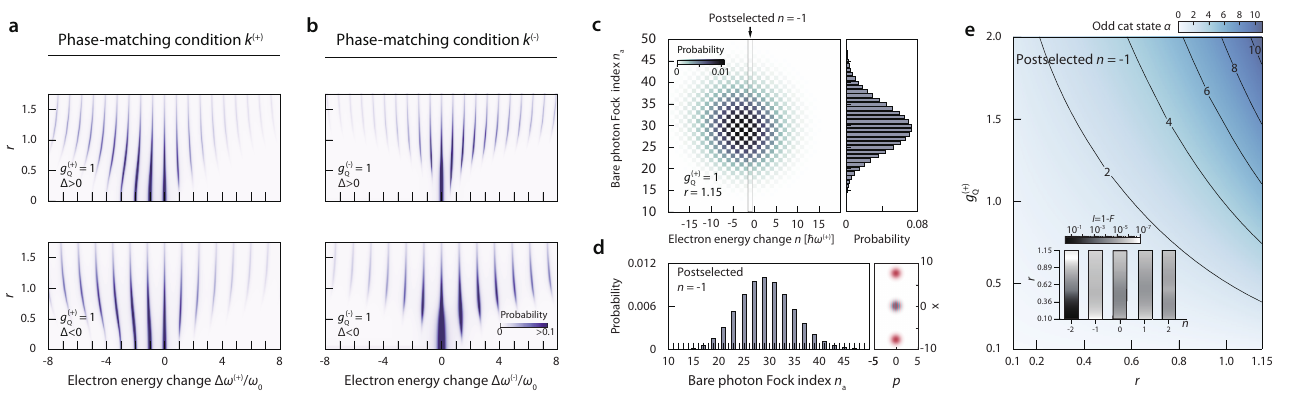}
	 \caption{%
	 	\textbf{Free-electron parametric interaction with bare vacuum and its generation of Schrödinger cat states.}
        %
	   \textbf{a-b.} Electron energy loss and gain spectra versus squeezing parameter $r$ for the two phase-matching conditions $\kplus$ (\textbf{a}) and $\kminus$ (\textbf{b}).
     The detuning is $\abs{\Delta}=0.2\omega_0$ and the Lorentzian broadening is $0.1\omega^{(\pm)}$, and electron energy change is normalized to $\omega_0$. 
        %
         \textbf{c.} 
         The joint probability distribution between the Fock state and the electron energy change $n \hbar \omegaplus$ for $\kplus$, exhibiting a checkerboard-like pattern.
         The right inset shows the corresponding traced near-Poissonian photon statistics.
        %
        %
         \textbf{d.} Postselection of even or odd loss/gain band enables the selection of even and odd photon Fock space (\eg, postselecting $n=-1$ in Fig.~(\textbf{c}) yields an odd-parity photon-number distribution), enabling the preparation of even or odd cat states.
        The right inset plots the Wigner function for the photonic state generated by postselecting $n=-1$.
        \textbf{e.} Cat amplitude $\alpha$ as a function of interaction strength $\gquplus$ and squeezing parameter $r$ for the $n=-1$ electron energy loss sideband.
        Inset: High fidelity of cat state preparation for various electron sideband heralding under different levels of squeezing parameter $r$.
        In all the plots, $\gquplus=1$. In (\textbf{a})-(\textbf{d}), $r=1.15$, and in (\textbf{a})-(\textbf{b}), the detuning is $\abs{\Delta}=0.2\omega_0$,  the Lorentzian broadening is $0.1\omega^{(\pm)}$, and electron energy change is normalized to $\omega_0$ . 
         }
	\label{fig_main:quantumsouruce_off}
\end{figure*}

Crucially, the phase matching of $\kplus$ and $\kminus$ \emph{only} corresponds to electron energy loss $n\hbar \omegaplus$ and gain $n\hbar \omegaminus$, respectively, giving rise to single-sided electron energy spectra with regard to the zero-loss peak [ZLP; Figs.~\ref{fig_main:quantumsouruce_on}(a)-\ref{fig_main:quantumsouruce_on}(b)].
In particular, under the phase-matching condition $\kplus$ [Fig.~\ref{fig_main:quantumsouruce_on}(a)], only energy loss $\Delta\omegaplus$ is observed in the electron energy spectrum, similar to the spontaneous electron energy loss at zero temperature. 
Intriguingly, for $\kminus$ [Fig.~\ref{fig_main:quantumsouruce_on}(b)], the spectrum exhibits exclusively energy gain sidebands.
This distinctive behavior can be harnessed to relax the phase-synchronization requirement of DLAs, which will be discussed in the following ``Quantum parametric dielectric laser accelerator" section.
Moreover, the sideband spacings $\hbar \omegaplus$ and $\hbar \omegaminus$ in the $\kplus$ and $\kminus$ phase-matching conditions can be either compressed ($\Delta>0$) or broadened ($\Delta<0$) as the squeezing parameter $r$ increases; both $\omegaplus$ and $\omegaminus$ approach $\omegap/2$ under diverging squeezing.

We calculate the joint probability to demonstrate the correlation between the electron energy change and the squeezed Fock state under the two phase-matching conditions $\kplus$ and $\kminus$, respectively [Figs.~\ref{fig_main:quantumsouruce_on}(c)-\ref{fig_main:quantumsouruce_on}(d);
see additional results for various interaction strengths and squeezing parameters in Sec. S3.B].
The two joint distributions clearly show a purely diagonal or anti-diagonal populations, indicating that conditioning on an electron energy loss of $n\hbar \omegaplus$ [for $\kplus$] or an energy gain of $n\hbar \omegaminus$ [for $\kminus$] projects the photonic mode uniquely onto a pure squeezed Fock state $\ket{n_\mathrm{s}=n}$.
The one-to-one entanglement between the electron energy state and the squeezed Fock state enables an electron energy measurement at the $-n$ (or $n$)-th sideband to herald the pure squeezed Fock state $\ket{n_\mathrm{s}}$.
By postselecting on the desired sideband, one can thereby achieve on-demand generation of squeezed Fock states.
This particular non-Gaussian state is predicted to serve as essential resources~\cite{tomoda2024boosting,korolev2024generation} for quantum information processing~\cite{kienzler2017quantum,bashmakova2025bosonic,cai2025quantum,zeng2025quantum} and quantum state engineering~\cite{winnel2024deterministic}.
Our approach provides a method for preparing arbitrary squeezed Fock states within the framework of free-electron quantum optics, along with the additional advantages for accessing high-order states brought by the quantum amplified interaction strength $\gplus$ or $\gminus$ [Eq.~\eqref{eq:g+&-}].

\subsection{Bare vacuum seeding for Schrödinger cat states}

We now switch off the quantum source in Fig.~\ref{fig_main:framework}(a) and study how the proposed setup interacts with the bare vacuum instead of the squeezed vacuum. 
The photonic state therefore starts from the bare vacuum $\ket{0_a}$, while the dynamics is still driven by the nonlinear DPDC through the undepleted pump.
It is natural to revert to the bare photon operators $\{ \hat{a}, \hat{a}^\dagger\}$ and rewrite the scattering matrices Eq.~\eqref{eq:S_bogo_two} in the bare photon basis $\ket{n_a}$ by substituting the inverse Bogoliubov transformation:
\begin{subequations}\label{eq:S_bare_two}
\begin{align}
&\begin{aligned}
\Scattbogoplus
 = & \mathrm{exp}
\Big\{  \cosh{r}
    \left[
    \cosh{r}\,{\gquplus}^{*}\, \bplus
        +\sinh{r}\, \gquplus \,\bplus{^\dagger}
         \right] \hat{a}^\dagger 
\\
&    - 
    \cosh{r} \left[ 
     \cosh{r}\, \gquplus \, \bplus{^\dagger}
    +\sinh{r} \, {\gquplus}^{*} \, \bplus
     \right] \hat{a}
\Big\}, 
\label{eq:S_bare_plus} 
\end{aligned}
\\
&\begin{aligned}
\Scattbogominus 
= &\mathrm{exp}
\Big\{ \sinh{r}
    \left[
     \sinh{r} \, {\gquminus} \,  \bminus{^\dagger}+ 
     \cosh{r}\, {\gquminus}^{*} \,\bminus
    \right] \hat{a} \,
\\ 
&   -
    \sinh{r} \left[
     \sinh{r} \, {\gquminus}^{*} \, \bminus +
      \cosh{r}\, \gquminus \,\bminus{^\dagger}
    \right] \hat{a}^\dagger
\Big\}.
\label{eq:S_bare_minus}
\end{aligned}
\end{align}
\end{subequations}
Here, because of the exponential scaling between photon number and Bogoliubov quasiparticle number, the interaction strength between the free electron and the bare photon is exponentially enhanced by $\sim [\exp(r)/2]^2$ ($r > 1$).
Eq.~\eqref{eq:S_bare_two} enables us to analyse the evolution in the bare photon basis and obtain the final state for a given initial electron--photon joint state $\ket{\Psi_{\mathrm i}}=\ket{E_0,0_a}$ (see Sec. S4.A).

Figures~\ref{fig_main:quantumsouruce_off}(a)-\ref{fig_main:quantumsouruce_off}(b) show the electron energy spectra for various squeezing parameters $r$ under the two phase-matching conditions $\kplus$ and $\kminus$, respectively.
The spectra exhibit both energy gain and loss sidebands in the absence of field injection at the interaction frequency. 
This contrasts the ``fieldless PINEM" (\ie, spontaneous EELS) with a bare vacuum $\ket{0_a}$, which yields only loss sidebands, and the parametric QPINEM seeded by a squeezed vacuum $\ket{0_\mathrm{s}}$, which yields only gain or only loss sidebands as discussed above. 
The simultaneous gain and loss sidebands arise because the bare optical modes $\ket{n_a}$ are not eigenmodes of the DPDC Hamiltonian.
Consequently, the bare vacuum already contains Bogoliubov quasiparticles, $\ket{0_a}=\Ssq(r) \ket{0_\mathrm{s}}$, allowing the electron to either absorb or emit such excitations.
Moreover, the spectra are asymmetric with respect to the ZLP for small $r$, because the electron energy ladder operators $\bplus$ [$\bminus$] and ${\bplus}{^\dagger}$ [${\bminus}{^\dagger}$] couple to the photon mode with unequal interaction strengths, as described in Eq.~\eqref{eq:S_bare_two}.
As $r$ increases, these interaction strengths gradually equalize and the spectra become symmetric about the ZLP.

We next turn to the photonic states using the phase-matching $\kplus$ Eq.~\eqref{eq:S_bare_plus} as an example; results for the phase-matching $\kminus$ Eq.~\eqref{eq:S_bare_minus} are given in the SM Sec. S4.C.
The joint electron--photon statistics in Fig.~\ref{fig_main:quantumsouruce_off}(c) reveal that bare photons can be generated regardless of whether the electron loses ($n<0$) or gains ($n>0$) energy.
The probability distribution acquires a pronounced “checkerboard’’ pattern, \ie, the probability oscillation in generating even or odd Fock states, which arises from the quadratic terms $\bplus{^2}$ and ${\bplus}{^{\dagger 2}}$ in the factorized scattering matrix (see Sec. S4.A).
The traced photon statistics, as shown in the right inset of Fig.~\ref{fig_main:quantumsouruce_off}(c), further indicate that the photon number distributions are nearly Poissonian (see the Poisson limits in Sec. S4.D).

The resulting joint statistics enable the preparation of cat states via electron postselection. 
The postselected photon statistics inherit the parity of the selected sideband: selecting an even or odd energy gain/loss projects the photonic state onto the even or odd Fock subspace [see the example in Fig.~\ref{fig_main:quantumsouruce_off}(d) for sideband $n = -1$].
Such parity-distinct projection suggests the generation of odd (even) cat states by postselecting the odd (even) sideband.
For example, postselecting the sideband $n = -1$ prepares an optical state with fidelity $F> 99\%$ to an odd cat state $\ket{\text{cat}_{\mathrm{o}}} \propto  \ket{\alpha} - \ket{-\alpha}$,
where the cat amplitude $\alpha$ increases with $r$ and $\gquplus$ [Fig.~\ref{fig_main:quantumsouruce_off}(e); see the fidelity calculation in SM Sec. S4.C]. 
The inset of Fig.~\ref{fig_main:quantumsouruce_off}(e) further demonstrates the high fidelity of state preparation for several sidebands at fixed $\gquplus = 1$. 
The same behaviour is obtained for the phase-matching $\kminus$ Eq.~\eqref{eq:S_bare_minus}; see the discussion in SM Sec. S4.C.

\subsection{Quantum parametric dielectric laser accelerators}

\begin{figure}[htbp]
	\centering
	\includegraphics[width=1\linewidth]{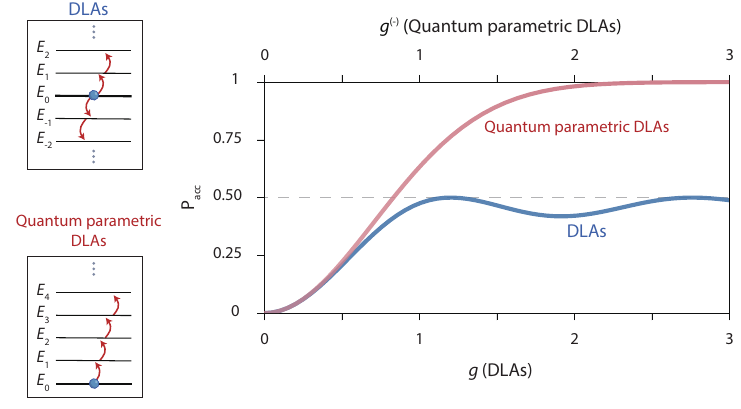}
	 \caption{%
	 	\textbf{Probability of electron acceleration in quantum parametric DLAs.} 
        %
        Insets: Electron energy-level diagrams for existing DLAs (upper-left) and the quantum parametric DLAs (lower-left), respectively.
        Right: Electron energy-gain probability $\mathrm{P}_{\mathrm{acc}}$ versus interaction strength $g$ for DLAs and $\gminus$ for the quantum parametric DLAs. 
        The probability of electron arrival is assumed to be uniformly distributed within an optical cycle.
        $\mathrm{P}_{\mathrm{acc}}$ saturates at 0.5 for DLAs, whereas it can reach unity for quantum parametric DLAs.
        }
	\label{fig_main:accelerator}
\end{figure}

As a unique application of the framework, we introduce quantum parametric DLAs, which mitigate the stringent requirement for precise temporal synchronization between optical field oscillations and electron arrival times---a fundamental constraint in DLAs. 
In the classical description of DLAs, net acceleration requires both momentum phase matching and temporal phase synchronization, \ie, the optical phase velocity equals the electron velocity so that the electron interacts with a constant field instead of an alternating field, and the arrival phase of the electron is required to synchronize with the accelerating half-cycle of the oscillating field.
If the arrival phases of electrons are uniformly distributed over one optical period, half of the electrons are accelerated, and half are decelerated, yielding zero ensemble-averaged energy change.
This half-half acceleration and deceleration result is consistent with the QPINEM description of DLAs:
when the electron coherence time far exceeds one optical period, a near-monochromatic coherent-state drive $\ket{\alpha}$ produces an electron energy spectrum with symmetric Bessel probability amplitudes about the ZLP, implying the ensemble-averaged energy exchange $\langle \Delta E \rangle =0$.
Thus, even after momentum phase matching is satisfied, net acceleration still requires temporal phase synchronization, which is achieved by shaping the electrons into microbunches much shorter than an optical cycle and phase-locking to the accelerating half-cycle of the optical field~\cite{shiloh2022miniature, peralta2013demonstration, sapra2020chip, shiloh2021electron,adiv2021quantum,chlouba2023coherent}.
Without these conditions, there can be no net acceleration in the quantum description.

The temporal synchronization requirement could be relaxed in the present framework by seeding the interaction with squeezed vacuum under the $\kminus$ phase-matching condition [Figs.~\ref{fig_main:quantumsouruce_on}(b) and~\ref{fig_main:quantumsouruce_on}(d)].
Under this configuration, electrons can only maintain their initial energy or gain energy quanta, as evidenced by the electron energy spectra of Fig.~\ref{fig_main:quantumsouruce_on}(b), which exhibit no loss sidebands.
To evaluate the probability of a randomly arrived electron being accelerated, we compare the energy gain probability $\mathrm{P}_{\mathrm{acc}}$ of the quantum parametric DLAs with that of existing DLAs [see the details in SM Sec. S5].
Figure~\ref{fig_main:accelerator} shows $\mathrm{P}_{\mathrm{acc}}$ as a function of the interaction strength $g=\alpha \gqu$ (DLAs) and $\gminus =\gquminus \, \sinh{r}$ (quantum parametric DLAs).
As $g$ and $\gminus$ increase, $\mathrm{P}_{\mathrm{acc}}$ of the existing DLAs saturates at~\SI{0.5}{} due to the equal probabilities for electrons to gain or lose energy. 
In contrast, the quantum parametric DLAs can reach $\mathrm{P}_{\mathrm{acc}}=\SI{1}{}$ for large interaction strengths, meaning that every injected electron can be accelerated regardless of its initial phase uncertainty regarding the optical cycle.

\subsection{Conclusion and outlook}

We introduced and developed a general framework for optical quantum parametric free-electron--photon interaction.
We showed that free electrons couple to Bogoliubov quasiparticles under a detuned drive, leading to a modified electron--photon phase-matching condition, an exponential enhancement of the coupling strength with squeezing, and qualitatively new electron-energy spectra. 
Leveraging these effects, we predict two complementary routes to non-Gaussian state generation, with distinct and directly measurable electron-energy spectral features and optical resources.
With a squeezed vacuum seed, the electron energy spectrum is loss-only or gain-only depending on the corresponding modified phase-matching condition, and the sideband index heralds a unique squeezed Fock state matched one-to-one to the electron’s energy change.
With a bare vacuum input, the spectrum simultaneously exhibits gain and loss sidebands, but both corresponding to the generation of photons.
Postselection on odd or even sidebands projects the photons onto high-fidelity odd or even Schrödinger cat states.
As an application, we introduced quantum parametric DLAs that relax temporal phase synchronisation and enable the accelerated-electron probability to approach unity for electrons uniformly distributed in optical phase.

Towards the experimental realization of the system, integrated on-chip nonlinear photonics~\cite{wang2020integrated, zhu2021integrated,dutt2024nonlinear} coupled with free electrons provides a highly promising platform.
In a cascaded configuration, a single-mode squeezed vacuum can be generated via a nonlinear waveguide and then coupled into a high-$Q$ nonlinear cavity interfacing free electrons.
\swap{The cavity should be designed to support either longitudinal wavevectors $\kplus$ or $\kminus$, and the detuned parametric drive is applied to the same mode to implement the intracavity DPDC process.
}
{The cavity should be designed such that its longitudinal mode wavevector matches either \(\kplus\) or \(\kminus\) at $\omega_0$, and the detuned parametric drive is applied to
the same mode to implement the intracavity DPDC process.
}
Alternatively, a single periodically poled nonlinear waveguide could potentially realize both squeezing generation and DPDC interaction.
In this scheme, the free electron can interact with the guided signal/idler mode over an extended length with large spatial overlap and benefit from the large coupling strength.
By pre-pumping the waveguide, a steady-state squeezed vacuum can be prepared for the electron--photon interaction; otherwise, the initial state remains the bare vacuum.

Our framework naturally extends to multimodal interactions, including nondegenerate parametric down-conversion and multimodal cavities:
the former can be treated via multimode Bogoliubov transformations, while the latter fits within a QPINEM description of electron coupling to multiple optical eigenmodes.
In multi-electron operation, successive electrons can become mutually entangled via the shared parametrically driven cavity, leading to cross-electron entanglement.
In addition, incorporating electron recoil and pondermotive light--matter coupling will further refine predictions for slow electrons and intense light fields.
Beyond the squeezed vacuum and bare vacuum seeding explored here, the initial photonic states can be generalized to other Gaussian or non-Gaussian states. 
Together with versatile electron wave-packet phase and momentum shaping, all these control knobs are expected to establish the parametric free-electron--photon interaction as an ideal platform for generating highly nontrivial electron and photon statistics.

\emph{Acknowledgments.}
We thank Di Zhu for fruitful discussions and suggestions.


%


\begin{thebibliography}{96}%
\makeatletter
\providecommand \@ifxundefined [1]{%
 \@ifx{#1\undefined}
}%
\providecommand \@ifnum [1]{%
 \ifnum #1\expandafter \@firstoftwo
 \else \expandafter \@secondoftwo
 \fi
}%
\providecommand \@ifx [1]{%
 \ifx #1\expandafter \@firstoftwo
 \else \expandafter \@secondoftwo
 \fi
}%
\providecommand \natexlab [1]{#1}%
\providecommand \enquote  [1]{``#1''}%
\providecommand \bibnamefont  [1]{#1}%
\providecommand \bibfnamefont [1]{#1}%
\providecommand \citenamefont [1]{#1}%
\providecommand \href@noop [0]{\@secondoftwo}%
\providecommand \href [0]{\begingroup \@sanitize@url \@href}%
\providecommand \@href[1]{\@@startlink{#1}\@@href}%
\providecommand \@@href[1]{\endgroup#1\@@endlink}%
\providecommand \@sanitize@url [0]{\catcode `\\12\catcode `\$12\catcode `\&12\catcode `\#12\catcode `\^12\catcode `\_12\catcode `\%12\relax}%
\providecommand \@@startlink[1]{}%
\providecommand \@@endlink[0]{}%
\providecommand \url  [0]{\begingroup\@sanitize@url \@url }%
\providecommand \@url [1]{\endgroup\@href {#1}{\urlprefix }}%
\providecommand \urlprefix  [0]{URL }%
\providecommand \Eprint [0]{\href }%
\providecommand \doibase [0]{https://doi.org/}%
\providecommand \selectlanguage [0]{\@gobble}%
\providecommand \bibinfo  [0]{\@secondoftwo}%
\providecommand \bibfield  [0]{\@secondoftwo}%
\providecommand \translation [1]{[#1]}%
\providecommand \BibitemOpen [0]{}%
\providecommand \bibitemStop [0]{}%
\providecommand \bibitemNoStop [0]{.\EOS\space}%
\providecommand \EOS [0]{\spacefactor3000\relax}%
\providecommand \BibitemShut  [1]{\csname bibitem#1\endcsname}%
\let\auto@bib@innerbib\@empty
\bibitem [{\citenamefont {Giordmaine}\ and\ \citenamefont {Miller}(1965)}]{giordmaine1965tunable}%
  \BibitemOpen
  \bibfield  {author} {\bibinfo {author} {\bibfnamefont {J.~A.}\ \bibnamefont {Giordmaine}}\ and\ \bibinfo {author} {\bibfnamefont {R.~C.}\ \bibnamefont {Miller}},\ }\href {https://doi.org/10.1103/PhysRevLett.14.973} {\bibfield  {journal} {\bibinfo  {journal} {Physical Review Letters}\ }\textbf {\bibinfo {volume} {14}},\ \bibinfo {pages} {973} (\bibinfo {year} {1965})}\BibitemShut {NoStop}%
\bibitem [{\citenamefont {Pepper}\ \emph {et~al.}(1978)\citenamefont {Pepper}, \citenamefont {Yariv},\ and\ \citenamefont {Fekete}}]{pepper1978observation}%
  \BibitemOpen
  \bibfield  {author} {\bibinfo {author} {\bibfnamefont {D.~M.}\ \bibnamefont {Pepper}}, \bibinfo {author} {\bibfnamefont {A.}~\bibnamefont {Yariv}},\ and\ \bibinfo {author} {\bibfnamefont {D.}~\bibnamefont {Fekete}},\ }\href {https://doi.org/10.1063/1.90185} {\bibfield  {journal} {\bibinfo  {journal} {Applied Physics Letters}\ }\textbf {\bibinfo {volume} {33}},\ \bibinfo {pages} {41} (\bibinfo {year} {1978})}\BibitemShut {NoStop}%
\bibitem [{\citenamefont {Wang}\ and\ \citenamefont {Racette}(1965)}]{wang1965measurement}%
  \BibitemOpen
  \bibfield  {author} {\bibinfo {author} {\bibfnamefont {C.~C.}\ \bibnamefont {Wang}}\ and\ \bibinfo {author} {\bibfnamefont {G.~W.}\ \bibnamefont {Racette}},\ }\href {https://doi.org/10.1063/1.1754219} {\bibfield  {journal} {\bibinfo  {journal} {Applied Physics Letters}\ }\textbf {\bibinfo {volume} {6}},\ \bibinfo {pages} {169} (\bibinfo {year} {1965})}\BibitemShut {NoStop}%
\bibitem [{\citenamefont {Carman}\ \emph {et~al.}(1966)\citenamefont {Carman}, \citenamefont {Chiao},\ and\ \citenamefont {Kelley}}]{carman1966observation}%
  \BibitemOpen
  \bibfield  {author} {\bibinfo {author} {\bibfnamefont {R.}~\bibnamefont {Carman}}, \bibinfo {author} {\bibfnamefont {R.}~\bibnamefont {Chiao}},\ and\ \bibinfo {author} {\bibfnamefont {P.}~\bibnamefont {Kelley}},\ }\href {https://doi.org/10.1103/PhysRevLett.17.1281} {\bibfield  {journal} {\bibinfo  {journal} {Physical Review Letters}\ }\textbf {\bibinfo {volume} {17}},\ \bibinfo {pages} {1281} (\bibinfo {year} {1966})}\BibitemShut {NoStop}%
\bibitem [{\citenamefont {Slusher}\ \emph {et~al.}(1985)\citenamefont {Slusher}, \citenamefont {Hollberg}, \citenamefont {Yurke}, \citenamefont {Mertz},\ and\ \citenamefont {Valley}}]{slusher1985observation}%
  \BibitemOpen
  \bibfield  {author} {\bibinfo {author} {\bibfnamefont {R.}~\bibnamefont {Slusher}}, \bibinfo {author} {\bibfnamefont {L.}~\bibnamefont {Hollberg}}, \bibinfo {author} {\bibfnamefont {B.}~\bibnamefont {Yurke}}, \bibinfo {author} {\bibfnamefont {J.}~\bibnamefont {Mertz}},\ and\ \bibinfo {author} {\bibfnamefont {J.}~\bibnamefont {Valley}},\ }\href {https://doi.org/10.1103/PhysRevLett.55.2409} {\bibfield  {journal} {\bibinfo  {journal} {Physical review letters}\ }\textbf {\bibinfo {volume} {55}},\ \bibinfo {pages} {2409} (\bibinfo {year} {1985})}\BibitemShut {NoStop}%
\bibitem [{\citenamefont {Wu}\ \emph {et~al.}(1986)\citenamefont {Wu}, \citenamefont {Kimble}, \citenamefont {Hall},\ and\ \citenamefont {Wu}}]{wu1986generation}%
  \BibitemOpen
  \bibfield  {author} {\bibinfo {author} {\bibfnamefont {L.-A.}\ \bibnamefont {Wu}}, \bibinfo {author} {\bibfnamefont {H.}~\bibnamefont {Kimble}}, \bibinfo {author} {\bibfnamefont {J.}~\bibnamefont {Hall}},\ and\ \bibinfo {author} {\bibfnamefont {H.}~\bibnamefont {Wu}},\ }\href {https://doi.org/10.1103/PhysRevLett.57.2520} {\bibfield  {journal} {\bibinfo  {journal} {Physical review letters}\ }\textbf {\bibinfo {volume} {57}},\ \bibinfo {pages} {2520} (\bibinfo {year} {1986})}\BibitemShut {NoStop}%
\bibitem [{\citenamefont {Shelby}\ \emph {et~al.}(1986)\citenamefont {Shelby}, \citenamefont {Levenson}, \citenamefont {Perlmutter}, \citenamefont {DeVoe},\ and\ \citenamefont {Walls}}]{shelby1986broad}%
  \BibitemOpen
  \bibfield  {author} {\bibinfo {author} {\bibfnamefont {R.~M.}\ \bibnamefont {Shelby}}, \bibinfo {author} {\bibfnamefont {M.~D.}\ \bibnamefont {Levenson}}, \bibinfo {author} {\bibfnamefont {S.~H.}\ \bibnamefont {Perlmutter}}, \bibinfo {author} {\bibfnamefont {R.~G.}\ \bibnamefont {DeVoe}},\ and\ \bibinfo {author} {\bibfnamefont {D.~F.}\ \bibnamefont {Walls}},\ }\href {https://doi.org/10.1103/PhysRevLett.57.691} {\bibfield  {journal} {\bibinfo  {journal} {Physical review letters}\ }\textbf {\bibinfo {volume} {57}},\ \bibinfo {pages} {691} (\bibinfo {year} {1986})}\BibitemShut {NoStop}%
\bibitem [{\citenamefont {Burnham}\ and\ \citenamefont {Weinberg}(1970)}]{burnham1970observation}%
  \BibitemOpen
  \bibfield  {author} {\bibinfo {author} {\bibfnamefont {D.~C.}\ \bibnamefont {Burnham}}\ and\ \bibinfo {author} {\bibfnamefont {D.~L.}\ \bibnamefont {Weinberg}},\ }\href {https://doi.org/10.1103/PhysRevLett.25.84} {\bibfield  {journal} {\bibinfo  {journal} {Physical Review Letters}\ }\textbf {\bibinfo {volume} {25}},\ \bibinfo {pages} {84} (\bibinfo {year} {1970})}\BibitemShut {NoStop}%
\bibitem [{\citenamefont {Kwiat}\ \emph {et~al.}(1995)\citenamefont {Kwiat}, \citenamefont {Mattle}, \citenamefont {Weinfurter}, \citenamefont {Zeilinger}, \citenamefont {Sergienko},\ and\ \citenamefont {Shih}}]{kwiat1995new}%
  \BibitemOpen
  \bibfield  {author} {\bibinfo {author} {\bibfnamefont {P.~G.}\ \bibnamefont {Kwiat}}, \bibinfo {author} {\bibfnamefont {K.}~\bibnamefont {Mattle}}, \bibinfo {author} {\bibfnamefont {H.}~\bibnamefont {Weinfurter}}, \bibinfo {author} {\bibfnamefont {A.}~\bibnamefont {Zeilinger}}, \bibinfo {author} {\bibfnamefont {A.~V.}\ \bibnamefont {Sergienko}},\ and\ \bibinfo {author} {\bibfnamefont {Y.}~\bibnamefont {Shih}},\ }\href {https://doi.org/10.1103/PhysRevLett.75.4337} {\bibfield  {journal} {\bibinfo  {journal} {Physical Review Letters}\ }\textbf {\bibinfo {volume} {75}},\ \bibinfo {pages} {4337} (\bibinfo {year} {1995})}\BibitemShut {NoStop}%
\bibitem [{\citenamefont {Wang}\ \emph {et~al.}(2020{\natexlab{a}})\citenamefont {Wang}, \citenamefont {Sciarrino}, \citenamefont {Laing},\ and\ \citenamefont {Thompson}}]{wang2020integrated}%
  \BibitemOpen
  \bibfield  {author} {\bibinfo {author} {\bibfnamefont {J.}~\bibnamefont {Wang}}, \bibinfo {author} {\bibfnamefont {F.}~\bibnamefont {Sciarrino}}, \bibinfo {author} {\bibfnamefont {A.}~\bibnamefont {Laing}},\ and\ \bibinfo {author} {\bibfnamefont {M.~G.}\ \bibnamefont {Thompson}},\ }\href {https://doi.org/10.1038/s41566-019-0532-1} {\bibfield  {journal} {\bibinfo  {journal} {Nature photonics}\ }\textbf {\bibinfo {volume} {14}},\ \bibinfo {pages} {273} (\bibinfo {year} {2020}{\natexlab{a}})}\BibitemShut {NoStop}%
\bibitem [{\citenamefont {Zhu}\ \emph {et~al.}(2021)\citenamefont {Zhu}, \citenamefont {Shao}, \citenamefont {Yu}, \citenamefont {Cheng}, \citenamefont {Desiatov}, \citenamefont {Xin}, \citenamefont {Hu}, \citenamefont {Holzgrafe}, \citenamefont {Ghosh}, \citenamefont {Shams-Ansari} \emph {et~al.}}]{zhu2021integrated}%
  \BibitemOpen
  \bibfield  {author} {\bibinfo {author} {\bibfnamefont {D.}~\bibnamefont {Zhu}}, \bibinfo {author} {\bibfnamefont {L.}~\bibnamefont {Shao}}, \bibinfo {author} {\bibfnamefont {M.}~\bibnamefont {Yu}}, \bibinfo {author} {\bibfnamefont {R.}~\bibnamefont {Cheng}}, \bibinfo {author} {\bibfnamefont {B.}~\bibnamefont {Desiatov}}, \bibinfo {author} {\bibfnamefont {C.~J.}\ \bibnamefont {Xin}}, \bibinfo {author} {\bibfnamefont {Y.}~\bibnamefont {Hu}}, \bibinfo {author} {\bibfnamefont {J.}~\bibnamefont {Holzgrafe}}, \bibinfo {author} {\bibfnamefont {S.}~\bibnamefont {Ghosh}}, \bibinfo {author} {\bibfnamefont {A.}~\bibnamefont {Shams-Ansari}}, \emph {et~al.},\ }\href {https://doi.org/10.1364/AOP.411024} {\bibfield  {journal} {\bibinfo  {journal} {Advances in Optics and Photonics}\ }\textbf {\bibinfo {volume} {13}},\ \bibinfo {pages} {242} (\bibinfo {year} {2021})}\BibitemShut {NoStop}%
\bibitem [{\citenamefont {Dutt}\ \emph {et~al.}(2024)\citenamefont {Dutt}, \citenamefont {Mohanty}, \citenamefont {Gaeta},\ and\ \citenamefont {Lipson}}]{dutt2024nonlinear}%
  \BibitemOpen
  \bibfield  {author} {\bibinfo {author} {\bibfnamefont {A.}~\bibnamefont {Dutt}}, \bibinfo {author} {\bibfnamefont {A.}~\bibnamefont {Mohanty}}, \bibinfo {author} {\bibfnamefont {A.~L.}\ \bibnamefont {Gaeta}},\ and\ \bibinfo {author} {\bibfnamefont {M.}~\bibnamefont {Lipson}},\ }\href {https://doi.org/10.1038/s41578-024-00668-z} {\bibfield  {journal} {\bibinfo  {journal} {Nature Reviews Materials}\ }\textbf {\bibinfo {volume} {9}},\ \bibinfo {pages} {321} (\bibinfo {year} {2024})}\BibitemShut {NoStop}%
\bibitem [{\citenamefont {L{\"u}}\ \emph {et~al.}(2015)\citenamefont {L{\"u}}, \citenamefont {Wu}, \citenamefont {Johansson}, \citenamefont {Jing}, \citenamefont {Zhang},\ and\ \citenamefont {Nori}}]{lu2015squeezed}%
  \BibitemOpen
  \bibfield  {author} {\bibinfo {author} {\bibfnamefont {X.-Y.}\ \bibnamefont {L{\"u}}}, \bibinfo {author} {\bibfnamefont {Y.}~\bibnamefont {Wu}}, \bibinfo {author} {\bibfnamefont {J.}~\bibnamefont {Johansson}}, \bibinfo {author} {\bibfnamefont {H.}~\bibnamefont {Jing}}, \bibinfo {author} {\bibfnamefont {J.}~\bibnamefont {Zhang}},\ and\ \bibinfo {author} {\bibfnamefont {F.}~\bibnamefont {Nori}},\ }\href {https://doi.org/10.1103/PhysRevLett.114.093602} {\bibfield  {journal} {\bibinfo  {journal} {Physical review letters}\ }\textbf {\bibinfo {volume} {114}},\ \bibinfo {pages} {093602} (\bibinfo {year} {2015})}\BibitemShut {NoStop}%
\bibitem [{\citenamefont {Zeytino{\u{g}}lu}\ \emph {et~al.}(2017)\citenamefont {Zeytino{\u{g}}lu}, \citenamefont {{\.I}mamo{\u{g}}lu},\ and\ \citenamefont {Huber}}]{zeytinouglu2017engineering}%
  \BibitemOpen
  \bibfield  {author} {\bibinfo {author} {\bibfnamefont {S.}~\bibnamefont {Zeytino{\u{g}}lu}}, \bibinfo {author} {\bibfnamefont {A.}~\bibnamefont {{\.I}mamo{\u{g}}lu}},\ and\ \bibinfo {author} {\bibfnamefont {S.}~\bibnamefont {Huber}},\ }\href {https://doi.org/10.1103/PhysRevX.7.021041} {\bibfield  {journal} {\bibinfo  {journal} {Physical Review X}\ }\textbf {\bibinfo {volume} {7}},\ \bibinfo {pages} {021041} (\bibinfo {year} {2017})}\BibitemShut {NoStop}%
\bibitem [{\citenamefont {Qin}\ \emph {et~al.}(2018)\citenamefont {Qin}, \citenamefont {Miranowicz}, \citenamefont {Li}, \citenamefont {L{\"u}}, \citenamefont {You},\ and\ \citenamefont {Nori}}]{qin2018exponentially}%
  \BibitemOpen
  \bibfield  {author} {\bibinfo {author} {\bibfnamefont {W.}~\bibnamefont {Qin}}, \bibinfo {author} {\bibfnamefont {A.}~\bibnamefont {Miranowicz}}, \bibinfo {author} {\bibfnamefont {P.-B.}\ \bibnamefont {Li}}, \bibinfo {author} {\bibfnamefont {X.-Y.}\ \bibnamefont {L{\"u}}}, \bibinfo {author} {\bibfnamefont {J.-Q.}\ \bibnamefont {You}},\ and\ \bibinfo {author} {\bibfnamefont {F.}~\bibnamefont {Nori}},\ }\href {https://doi.org/10.1103/PhysRevLett.120.093601} {\bibfield  {journal} {\bibinfo  {journal} {Physical Review Letters}\ }\textbf {\bibinfo {volume} {120}},\ \bibinfo {pages} {093601} (\bibinfo {year} {2018})}\BibitemShut {NoStop}%
\bibitem [{\citenamefont {S{\'a}nchez~Mu{\~n}oz}\ and\ \citenamefont {Jaksch}(2021)}]{sanchez2021squeezed}%
  \BibitemOpen
  \bibfield  {author} {\bibinfo {author} {\bibfnamefont {C.}~\bibnamefont {S{\'a}nchez~Mu{\~n}oz}}\ and\ \bibinfo {author} {\bibfnamefont {D.}~\bibnamefont {Jaksch}},\ }\href {https://doi.org/10.1103/PhysRevLett.127.183603} {\bibfield  {journal} {\bibinfo  {journal} {Physical Review Letters}\ }\textbf {\bibinfo {volume} {127}},\ \bibinfo {pages} {183603} (\bibinfo {year} {2021})}\BibitemShut {NoStop}%
\bibitem [{\citenamefont {L{\^e}}\ \emph {et~al.}(2025)\citenamefont {L{\^e}}, \citenamefont {Lukin}, \citenamefont {Roques-Carmes}, \citenamefont {Karnieli}, \citenamefont {Lustig}, \citenamefont {Guidry}, \citenamefont {Fan},\ and\ \citenamefont {Vu{\v{c}}kovi{\'c}}}]{le2025cavity}%
  \BibitemOpen
  \bibfield  {author} {\bibinfo {author} {\bibfnamefont {T.~K.}\ \bibnamefont {L{\^e}}}, \bibinfo {author} {\bibfnamefont {D.~M.}\ \bibnamefont {Lukin}}, \bibinfo {author} {\bibfnamefont {C.}~\bibnamefont {Roques-Carmes}}, \bibinfo {author} {\bibfnamefont {A.}~\bibnamefont {Karnieli}}, \bibinfo {author} {\bibfnamefont {E.}~\bibnamefont {Lustig}}, \bibinfo {author} {\bibfnamefont {M.~A.}\ \bibnamefont {Guidry}}, \bibinfo {author} {\bibfnamefont {S.}~\bibnamefont {Fan}},\ and\ \bibinfo {author} {\bibfnamefont {J.}~\bibnamefont {Vu{\v{c}}kovi{\'c}}},\ }\href {https://doi.org/10.1103/8qtt-symt} {\bibfield  {journal} {\bibinfo  {journal} {Physical Review Applied}\ }\textbf {\bibinfo {volume} {24}},\ \bibinfo {pages} {034053} (\bibinfo {year} {2025})}\BibitemShut {NoStop}%
\bibitem [{\citenamefont {Forn-D{\'\i}az}\ \emph {et~al.}(2019)\citenamefont {Forn-D{\'\i}az}, \citenamefont {Lamata}, \citenamefont {Rico}, \citenamefont {Kono},\ and\ \citenamefont {Solano}}]{forn2019ultrastrong}%
  \BibitemOpen
  \bibfield  {author} {\bibinfo {author} {\bibfnamefont {P.}~\bibnamefont {Forn-D{\'\i}az}}, \bibinfo {author} {\bibfnamefont {L.}~\bibnamefont {Lamata}}, \bibinfo {author} {\bibfnamefont {E.}~\bibnamefont {Rico}}, \bibinfo {author} {\bibfnamefont {J.}~\bibnamefont {Kono}},\ and\ \bibinfo {author} {\bibfnamefont {E.}~\bibnamefont {Solano}},\ }\href {https://doi.org/10.1103/RevModPhys.91.025005} {\bibfield  {journal} {\bibinfo  {journal} {Reviews of Modern Physics}\ }\textbf {\bibinfo {volume} {91}},\ \bibinfo {pages} {025005} (\bibinfo {year} {2019})}\BibitemShut {NoStop}%
\bibitem [{\citenamefont {Qin}\ \emph {et~al.}(2024)\citenamefont {Qin}, \citenamefont {Kockum}, \citenamefont {Mu{\~n}oz}, \citenamefont {Miranowicz},\ and\ \citenamefont {Nori}}]{qin2024quantum}%
  \BibitemOpen
  \bibfield  {author} {\bibinfo {author} {\bibfnamefont {W.}~\bibnamefont {Qin}}, \bibinfo {author} {\bibfnamefont {A.~F.}\ \bibnamefont {Kockum}}, \bibinfo {author} {\bibfnamefont {C.~S.}\ \bibnamefont {Mu{\~n}oz}}, \bibinfo {author} {\bibfnamefont {A.}~\bibnamefont {Miranowicz}},\ and\ \bibinfo {author} {\bibfnamefont {F.}~\bibnamefont {Nori}},\ }\href {https://doi.org/10.1016/j.physrep.2024.05.003} {\bibfield  {journal} {\bibinfo  {journal} {Physics Reports}\ }\textbf {\bibinfo {volume} {1078}},\ \bibinfo {pages} {1} (\bibinfo {year} {2024})}\BibitemShut {NoStop}%
\bibitem [{\citenamefont {Cerenkov}(1934)}]{cerenkov1934visible}%
  \BibitemOpen
  \bibfield  {author} {\bibinfo {author} {\bibfnamefont {P.}~\bibnamefont {Cerenkov}},\ }in\ \href@noop {} {\emph {\bibinfo {booktitle} {Dokl. Akad. Nauk SSSR}}},\ Vol.~\bibinfo {volume} {2}\ (\bibinfo {year} {1934})\ p.\ \bibinfo {pages} {451}\BibitemShut {NoStop}%
\bibitem [{\citenamefont {Friedman}\ \emph {et~al.}(1988)\citenamefont {Friedman}, \citenamefont {Gover}, \citenamefont {Kurizki}, \citenamefont {Ruschin},\ and\ \citenamefont {Yariv}}]{friedman1988spontaneous}%
  \BibitemOpen
  \bibfield  {author} {\bibinfo {author} {\bibfnamefont {A.}~\bibnamefont {Friedman}}, \bibinfo {author} {\bibfnamefont {A.}~\bibnamefont {Gover}}, \bibinfo {author} {\bibfnamefont {G.}~\bibnamefont {Kurizki}}, \bibinfo {author} {\bibfnamefont {S.}~\bibnamefont {Ruschin}},\ and\ \bibinfo {author} {\bibfnamefont {A.}~\bibnamefont {Yariv}},\ }\href {https://doi.org/10.1103/RevModPhys.60.471} {\bibfield  {journal} {\bibinfo  {journal} {Reviews of modern physics}\ }\textbf {\bibinfo {volume} {60}},\ \bibinfo {pages} {471} (\bibinfo {year} {1988})}\BibitemShut {NoStop}%
\bibitem [{\citenamefont {Garc{\'\i}a~de Abajo}(2010)}]{garcia2010optical}%
  \BibitemOpen
  \bibfield  {author} {\bibinfo {author} {\bibfnamefont {F.~J.}\ \bibnamefont {Garc{\'\i}a~de Abajo}},\ }\href {https://doi.org/10.1103/RevModPhys.82.209} {\bibfield  {journal} {\bibinfo  {journal} {Reviews of modern physics}\ }\textbf {\bibinfo {volume} {82}},\ \bibinfo {pages} {209} (\bibinfo {year} {2010})}\BibitemShut {NoStop}%
\bibitem [{\citenamefont {Liu}\ \emph {et~al.}(2017)\citenamefont {Liu}, \citenamefont {Xiao}, \citenamefont {Ye}, \citenamefont {Wang}, \citenamefont {Cui}, \citenamefont {Feng}, \citenamefont {Zhang},\ and\ \citenamefont {Huang}}]{liu2017integrated}%
  \BibitemOpen
  \bibfield  {author} {\bibinfo {author} {\bibfnamefont {F.}~\bibnamefont {Liu}}, \bibinfo {author} {\bibfnamefont {L.}~\bibnamefont {Xiao}}, \bibinfo {author} {\bibfnamefont {Y.}~\bibnamefont {Ye}}, \bibinfo {author} {\bibfnamefont {M.}~\bibnamefont {Wang}}, \bibinfo {author} {\bibfnamefont {K.}~\bibnamefont {Cui}}, \bibinfo {author} {\bibfnamefont {X.}~\bibnamefont {Feng}}, \bibinfo {author} {\bibfnamefont {W.}~\bibnamefont {Zhang}},\ and\ \bibinfo {author} {\bibfnamefont {Y.}~\bibnamefont {Huang}},\ }\href {https://doi.org/10.1038/nphoton.2017.45} {\bibfield  {journal} {\bibinfo  {journal} {Nature Photonics}\ }\textbf {\bibinfo {volume} {11}},\ \bibinfo {pages} {289} (\bibinfo {year} {2017})}\BibitemShut {NoStop}%
\bibitem [{\citenamefont {Polman}\ \emph {et~al.}(2019)\citenamefont {Polman}, \citenamefont {Kociak},\ and\ \citenamefont {Garc{\'\i}a~de Abajo}}]{polman2019electron}%
  \BibitemOpen
  \bibfield  {author} {\bibinfo {author} {\bibfnamefont {A.}~\bibnamefont {Polman}}, \bibinfo {author} {\bibfnamefont {M.}~\bibnamefont {Kociak}},\ and\ \bibinfo {author} {\bibfnamefont {F.~J.}\ \bibnamefont {Garc{\'\i}a~de Abajo}},\ }\href {https://doi.org/10.1038/s41563-019-0409-1} {\bibfield  {journal} {\bibinfo  {journal} {Nature materials}\ }\textbf {\bibinfo {volume} {18}},\ \bibinfo {pages} {1158} (\bibinfo {year} {2019})}\BibitemShut {NoStop}%
\bibitem [{\citenamefont {Shiloh}\ \emph {et~al.}(2022)\citenamefont {Shiloh}, \citenamefont {Sch{\"o}nenberger}, \citenamefont {Adiv}, \citenamefont {Ruimy}, \citenamefont {Karnieli}, \citenamefont {Hughes}, \citenamefont {Joel~England}, \citenamefont {Leedle}, \citenamefont {Black}, \citenamefont {Zhao} \emph {et~al.}}]{shiloh2022miniature}%
  \BibitemOpen
  \bibfield  {author} {\bibinfo {author} {\bibfnamefont {R.}~\bibnamefont {Shiloh}}, \bibinfo {author} {\bibfnamefont {N.}~\bibnamefont {Sch{\"o}nenberger}}, \bibinfo {author} {\bibfnamefont {Y.}~\bibnamefont {Adiv}}, \bibinfo {author} {\bibfnamefont {R.}~\bibnamefont {Ruimy}}, \bibinfo {author} {\bibfnamefont {A.}~\bibnamefont {Karnieli}}, \bibinfo {author} {\bibfnamefont {T.}~\bibnamefont {Hughes}}, \bibinfo {author} {\bibfnamefont {R.}~\bibnamefont {Joel~England}}, \bibinfo {author} {\bibfnamefont {K.~J.}\ \bibnamefont {Leedle}}, \bibinfo {author} {\bibfnamefont {D.~S.}\ \bibnamefont {Black}}, \bibinfo {author} {\bibfnamefont {Z.}~\bibnamefont {Zhao}}, \emph {et~al.},\ }\href {https://doi.org/10.1364/AOP.461142} {\bibfield  {journal} {\bibinfo  {journal} {Advances in Optics and Photonics}\ }\textbf {\bibinfo {volume} {14}},\ \bibinfo {pages} {862} (\bibinfo {year} {2022})}\BibitemShut {NoStop}%
\bibitem [{\citenamefont {Yang}\ \emph {et~al.}(2023)\citenamefont {Yang}, \citenamefont {Roques-Carmes}, \citenamefont {Kooi}, \citenamefont {Tang}, \citenamefont {Beroz}, \citenamefont {Mazur}, \citenamefont {Kaminer}, \citenamefont {Joannopoulos},\ and\ \citenamefont {Solja{\v{c}}i{\'c}}}]{yang2023photonic}%
  \BibitemOpen
  \bibfield  {author} {\bibinfo {author} {\bibfnamefont {Y.}~\bibnamefont {Yang}}, \bibinfo {author} {\bibfnamefont {C.}~\bibnamefont {Roques-Carmes}}, \bibinfo {author} {\bibfnamefont {S.~E.}\ \bibnamefont {Kooi}}, \bibinfo {author} {\bibfnamefont {H.}~\bibnamefont {Tang}}, \bibinfo {author} {\bibfnamefont {J.}~\bibnamefont {Beroz}}, \bibinfo {author} {\bibfnamefont {E.}~\bibnamefont {Mazur}}, \bibinfo {author} {\bibfnamefont {I.}~\bibnamefont {Kaminer}}, \bibinfo {author} {\bibfnamefont {J.~D.}\ \bibnamefont {Joannopoulos}},\ and\ \bibinfo {author} {\bibfnamefont {M.}~\bibnamefont {Solja{\v{c}}i{\'c}}},\ }\href {https://doi.org/10.1038/s41586-022-05387-5} {\bibfield  {journal} {\bibinfo  {journal} {Nature}\ }\textbf {\bibinfo {volume} {613}},\ \bibinfo {pages} {42} (\bibinfo {year} {2023})}\BibitemShut {NoStop}%
\bibitem [{\citenamefont {Roques-Carmes}\ \emph {et~al.}(2022)\citenamefont {Roques-Carmes}, \citenamefont {Rivera}, \citenamefont {Ghorashi}, \citenamefont {Kooi}, \citenamefont {Yang}, \citenamefont {Lin}, \citenamefont {Beroz}, \citenamefont {Massuda}, \citenamefont {Sloan}, \citenamefont {Romeo} \emph {et~al.}}]{roques2022framework}%
  \BibitemOpen
  \bibfield  {author} {\bibinfo {author} {\bibfnamefont {C.}~\bibnamefont {Roques-Carmes}}, \bibinfo {author} {\bibfnamefont {N.}~\bibnamefont {Rivera}}, \bibinfo {author} {\bibfnamefont {A.}~\bibnamefont {Ghorashi}}, \bibinfo {author} {\bibfnamefont {S.~E.}\ \bibnamefont {Kooi}}, \bibinfo {author} {\bibfnamefont {Y.}~\bibnamefont {Yang}}, \bibinfo {author} {\bibfnamefont {Z.}~\bibnamefont {Lin}}, \bibinfo {author} {\bibfnamefont {J.}~\bibnamefont {Beroz}}, \bibinfo {author} {\bibfnamefont {A.}~\bibnamefont {Massuda}}, \bibinfo {author} {\bibfnamefont {J.}~\bibnamefont {Sloan}}, \bibinfo {author} {\bibfnamefont {N.}~\bibnamefont {Romeo}}, \emph {et~al.},\ }\href {https://doi.org/10.1126/science.abm9293} {\bibfield  {journal} {\bibinfo  {journal} {Science}\ }\textbf {\bibinfo {volume} {375}},\ \bibinfo {pages} {eabm9293} (\bibinfo {year} {2022})}\BibitemShut {NoStop}%
\bibitem [{\citenamefont {Roques-Carmes}\ \emph {et~al.}(2023)\citenamefont {Roques-Carmes}, \citenamefont {Kooi}, \citenamefont {Yang}, \citenamefont {Rivera}, \citenamefont {Keathley}, \citenamefont {Joannopoulos}, \citenamefont {Johnson}, \citenamefont {Kaminer}, \citenamefont {Berggren},\ and\ \citenamefont {Soljačić}}]{roques2023free}%
  \BibitemOpen
  \bibfield  {author} {\bibinfo {author} {\bibfnamefont {C.}~\bibnamefont {Roques-Carmes}}, \bibinfo {author} {\bibfnamefont {S.~E.}\ \bibnamefont {Kooi}}, \bibinfo {author} {\bibfnamefont {Y.}~\bibnamefont {Yang}}, \bibinfo {author} {\bibfnamefont {N.}~\bibnamefont {Rivera}}, \bibinfo {author} {\bibfnamefont {P.~D.}\ \bibnamefont {Keathley}}, \bibinfo {author} {\bibfnamefont {J.~D.}\ \bibnamefont {Joannopoulos}}, \bibinfo {author} {\bibfnamefont {S.~G.}\ \bibnamefont {Johnson}}, \bibinfo {author} {\bibfnamefont {I.}~\bibnamefont {Kaminer}}, \bibinfo {author} {\bibfnamefont {K.~K.}\ \bibnamefont {Berggren}},\ and\ \bibinfo {author} {\bibfnamefont {M.}~\bibnamefont {Soljačić}},\ }\href {https://doi.org/10.1063/5.0118096} {\bibfield  {journal} {\bibinfo  {journal} {Applied Physics Reviews}\ }\textbf {\bibinfo {volume} {10}},\ \bibinfo {pages} {011303} (\bibinfo {year} {2023})}\BibitemShut {NoStop}%
\bibitem [{\citenamefont {Shi}\ \emph {et~al.}(2025)\citenamefont {Shi}, \citenamefont {Lee}, \citenamefont {Karnieli}, \citenamefont {Lohse}, \citenamefont {Gorlach}, \citenamefont {Wong}, \citenamefont {Salditt}, \citenamefont {Fan}, \citenamefont {Kaminer},\ and\ \citenamefont {Wong}}]{shi2025quantum}%
  \BibitemOpen
  \bibfield  {author} {\bibinfo {author} {\bibfnamefont {X.}~\bibnamefont {Shi}}, \bibinfo {author} {\bibfnamefont {W.~W.}\ \bibnamefont {Lee}}, \bibinfo {author} {\bibfnamefont {A.}~\bibnamefont {Karnieli}}, \bibinfo {author} {\bibfnamefont {L.~M.}\ \bibnamefont {Lohse}}, \bibinfo {author} {\bibfnamefont {A.}~\bibnamefont {Gorlach}}, \bibinfo {author} {\bibfnamefont {L.~W.~W.}\ \bibnamefont {Wong}}, \bibinfo {author} {\bibfnamefont {T.}~\bibnamefont {Salditt}}, \bibinfo {author} {\bibfnamefont {S.}~\bibnamefont {Fan}}, \bibinfo {author} {\bibfnamefont {I.}~\bibnamefont {Kaminer}},\ and\ \bibinfo {author} {\bibfnamefont {L.~J.}\ \bibnamefont {Wong}},\ }\href {https://doi.org/10.1016/j.pquantelec.2025.100577} {\bibfield  {journal} {\bibinfo  {journal} {Progress in Quantum Electronics}\ ,\ \bibinfo {pages} {100577}} (\bibinfo {year} {2025})}\BibitemShut {NoStop}%
\bibitem [{\citenamefont {Barwick}\ \emph {et~al.}(2009)\citenamefont {Barwick}, \citenamefont {Flannigan},\ and\ \citenamefont {Zewail}}]{barwick2009photon}%
  \BibitemOpen
  \bibfield  {author} {\bibinfo {author} {\bibfnamefont {B.}~\bibnamefont {Barwick}}, \bibinfo {author} {\bibfnamefont {D.~J.}\ \bibnamefont {Flannigan}},\ and\ \bibinfo {author} {\bibfnamefont {A.~H.}\ \bibnamefont {Zewail}},\ }\href {https://doi.org/10.1038/nature08662} {\bibfield  {journal} {\bibinfo  {journal} {Nature}\ }\textbf {\bibinfo {volume} {462}},\ \bibinfo {pages} {902} (\bibinfo {year} {2009})}\BibitemShut {NoStop}%
\bibitem [{\citenamefont {Garc{\'\i}a~de Abajo}\ \emph {et~al.}(2010)\citenamefont {Garc{\'\i}a~de Abajo}, \citenamefont {Asenjo-Garcia},\ and\ \citenamefont {Kociak}}]{garcia2010multiphoton}%
  \BibitemOpen
  \bibfield  {author} {\bibinfo {author} {\bibfnamefont {F.~J.}\ \bibnamefont {Garc{\'\i}a~de Abajo}}, \bibinfo {author} {\bibfnamefont {A.}~\bibnamefont {Asenjo-Garcia}},\ and\ \bibinfo {author} {\bibfnamefont {M.}~\bibnamefont {Kociak}},\ }\href {https://doi.org/10.1021/nl100613s} {\bibfield  {journal} {\bibinfo  {journal} {Nano letters}\ }\textbf {\bibinfo {volume} {10}},\ \bibinfo {pages} {1859} (\bibinfo {year} {2010})}\BibitemShut {NoStop}%
\bibitem [{\citenamefont {Park}\ \emph {et~al.}(2010)\citenamefont {Park}, \citenamefont {Lin},\ and\ \citenamefont {Zewail}}]{park2010photon}%
  \BibitemOpen
  \bibfield  {author} {\bibinfo {author} {\bibfnamefont {S.~T.}\ \bibnamefont {Park}}, \bibinfo {author} {\bibfnamefont {M.}~\bibnamefont {Lin}},\ and\ \bibinfo {author} {\bibfnamefont {A.~H.}\ \bibnamefont {Zewail}},\ }\href {https://doi.org/10.1088/1367-2630/12/12/123028} {\bibfield  {journal} {\bibinfo  {journal} {New Journal of Physics}\ }\textbf {\bibinfo {volume} {12}},\ \bibinfo {pages} {123028} (\bibinfo {year} {2010})}\BibitemShut {NoStop}%
\bibitem [{\citenamefont {Talebi}(2018)}]{talebi2018electron}%
  \BibitemOpen
  \bibfield  {author} {\bibinfo {author} {\bibfnamefont {N.}~\bibnamefont {Talebi}},\ }\href {https://doi.org/10.1080/23746149.2018.1499438} {\bibfield  {journal} {\bibinfo  {journal} {Advances in Physics: X}\ }\textbf {\bibinfo {volume} {3}},\ \bibinfo {pages} {1499438} (\bibinfo {year} {2018})}\BibitemShut {NoStop}%
\bibitem [{\citenamefont {Rivera}\ and\ \citenamefont {Kaminer}(2020)}]{rivera2020light}%
  \BibitemOpen
  \bibfield  {author} {\bibinfo {author} {\bibfnamefont {N.}~\bibnamefont {Rivera}}\ and\ \bibinfo {author} {\bibfnamefont {I.}~\bibnamefont {Kaminer}},\ }\href {https://doi.org/10.1038/s42254-020-0224-2} {\bibfield  {journal} {\bibinfo  {journal} {Nature Reviews Physics}\ }\textbf {\bibinfo {volume} {2}},\ \bibinfo {pages} {538} (\bibinfo {year} {2020})}\BibitemShut {NoStop}%
\bibitem [{\citenamefont {Garc{\'\i}a~de Abajo}\ and\ \citenamefont {Di~Giulio}(2021)}]{garcia2021optical}%
  \BibitemOpen
  \bibfield  {author} {\bibinfo {author} {\bibfnamefont {F.~J.}\ \bibnamefont {Garc{\'\i}a~de Abajo}}\ and\ \bibinfo {author} {\bibfnamefont {V.}~\bibnamefont {Di~Giulio}},\ }\href {https://doi.org/10.1021/acsphotonics.0c01950} {\bibfield  {journal} {\bibinfo  {journal} {ACS photonics}\ }\textbf {\bibinfo {volume} {8}},\ \bibinfo {pages} {945} (\bibinfo {year} {2021})}\BibitemShut {NoStop}%
\bibitem [{\citenamefont {Ruimy}\ \emph {et~al.}(2025)\citenamefont {Ruimy}, \citenamefont {Karnieli},\ and\ \citenamefont {Kaminer}}]{ruimy2025free}%
  \BibitemOpen
  \bibfield  {author} {\bibinfo {author} {\bibfnamefont {R.}~\bibnamefont {Ruimy}}, \bibinfo {author} {\bibfnamefont {A.}~\bibnamefont {Karnieli}},\ and\ \bibinfo {author} {\bibfnamefont {I.}~\bibnamefont {Kaminer}},\ }\href {https://doi.org/10.1038/s41567-024-02743-2} {\bibfield  {journal} {\bibinfo  {journal} {Nature Physics}\ }\textbf {\bibinfo {volume} {21}},\ \bibinfo {pages} {193} (\bibinfo {year} {2025})}\BibitemShut {NoStop}%
\bibitem [{\citenamefont {de~Abajo}\ \emph {et~al.}(2025)\citenamefont {de~Abajo}, \citenamefont {Polman}, \citenamefont {Velasco}, \citenamefont {Kociak}, \citenamefont {Tizei}, \citenamefont {St{\'e}phan}, \citenamefont {Meuret}, \citenamefont {Sannomiya}, \citenamefont {Akiba}, \citenamefont {Auad} \emph {et~al.}}]{garcia2025roadmap}%
  \BibitemOpen
  \bibfield  {author} {\bibinfo {author} {\bibfnamefont {F.~J.~G.}\ \bibnamefont {de~Abajo}}, \bibinfo {author} {\bibfnamefont {A.}~\bibnamefont {Polman}}, \bibinfo {author} {\bibfnamefont {C.~I.}\ \bibnamefont {Velasco}}, \bibinfo {author} {\bibfnamefont {M.}~\bibnamefont {Kociak}}, \bibinfo {author} {\bibfnamefont {L.~H.}\ \bibnamefont {Tizei}}, \bibinfo {author} {\bibfnamefont {O.}~\bibnamefont {St{\'e}phan}}, \bibinfo {author} {\bibfnamefont {S.}~\bibnamefont {Meuret}}, \bibinfo {author} {\bibfnamefont {T.}~\bibnamefont {Sannomiya}}, \bibinfo {author} {\bibfnamefont {K.}~\bibnamefont {Akiba}}, \bibinfo {author} {\bibfnamefont {Y.}~\bibnamefont {Auad}}, \emph {et~al.},\ }\href {https://doi.org/10.1021/acsphotonics.5c00585} {\bibfield  {journal} {\bibinfo  {journal} {ACS photonics}\ }\textbf {\bibinfo {volume} {12}},\ \bibinfo {pages} {4760} (\bibinfo {year} {2025})}\BibitemShut {NoStop}%
\bibitem [{\citenamefont {Kfir}(2019)}]{kfir2019entanglements}%
  \BibitemOpen
  \bibfield  {author} {\bibinfo {author} {\bibfnamefont {O.}~\bibnamefont {Kfir}},\ }\href {https://doi.org/10.1103/PhysRevLett.123.103602} {\bibfield  {journal} {\bibinfo  {journal} {Physical Review Letters}\ }\textbf {\bibinfo {volume} {123}},\ \bibinfo {pages} {103602} (\bibinfo {year} {2019})}\BibitemShut {NoStop}%
\bibitem [{\citenamefont {Di~Giulio}\ \emph {et~al.}(2019)\citenamefont {Di~Giulio}, \citenamefont {Kociak},\ and\ \citenamefont {Garc{\'\i}a~de Abajo}}]{di2019probing}%
  \BibitemOpen
  \bibfield  {author} {\bibinfo {author} {\bibfnamefont {V.}~\bibnamefont {Di~Giulio}}, \bibinfo {author} {\bibfnamefont {M.}~\bibnamefont {Kociak}},\ and\ \bibinfo {author} {\bibfnamefont {F.~J.}\ \bibnamefont {Garc{\'\i}a~de Abajo}},\ }\href {https://doi.org/10.1364/OPTICA.6.001524} {\bibfield  {journal} {\bibinfo  {journal} {Optica}\ }\textbf {\bibinfo {volume} {6}},\ \bibinfo {pages} {1524} (\bibinfo {year} {2019})}\BibitemShut {NoStop}%
\bibitem [{\citenamefont {Ben~Hayun}\ \emph {et~al.}(2021)\citenamefont {Ben~Hayun}, \citenamefont {Reinhardt}, \citenamefont {Nemirovsky}, \citenamefont {Karnieli}, \citenamefont {Rivera},\ and\ \citenamefont {Kaminer}}]{ben2021shaping}%
  \BibitemOpen
  \bibfield  {author} {\bibinfo {author} {\bibfnamefont {A.}~\bibnamefont {Ben~Hayun}}, \bibinfo {author} {\bibfnamefont {O.}~\bibnamefont {Reinhardt}}, \bibinfo {author} {\bibfnamefont {J.}~\bibnamefont {Nemirovsky}}, \bibinfo {author} {\bibfnamefont {A.}~\bibnamefont {Karnieli}}, \bibinfo {author} {\bibfnamefont {N.}~\bibnamefont {Rivera}},\ and\ \bibinfo {author} {\bibfnamefont {I.}~\bibnamefont {Kaminer}},\ }\href {https://doi.org/10.1126/sciadv.abe4270} {\bibfield  {journal} {\bibinfo  {journal} {Science Advances}\ }\textbf {\bibinfo {volume} {7}},\ \bibinfo {pages} {eabe4270} (\bibinfo {year} {2021})}\BibitemShut {NoStop}%
\bibitem [{\citenamefont {Henke}\ \emph {et~al.}(2025)\citenamefont {Henke}, \citenamefont {Jeng}, \citenamefont {Sivis},\ and\ \citenamefont {Ropers}}]{henke2025observation}%
  \BibitemOpen
  \bibfield  {author} {\bibinfo {author} {\bibfnamefont {J.-W.}\ \bibnamefont {Henke}}, \bibinfo {author} {\bibfnamefont {H.}~\bibnamefont {Jeng}}, \bibinfo {author} {\bibfnamefont {M.}~\bibnamefont {Sivis}},\ and\ \bibinfo {author} {\bibfnamefont {C.}~\bibnamefont {Ropers}},\ }\bibfield  {journal} {\bibinfo  {journal} {arXiv preprint arXiv:2504.13047}\ }\href {https://doi.org/10.48550/arXiv.2504.13047} {10.48550/arXiv.2504.13047} (\bibinfo {year} {2025})\BibitemShut {NoStop}%
\bibitem [{\citenamefont {Preimesberger}\ \emph {et~al.}(2025)\citenamefont {Preimesberger}, \citenamefont {Bogdanov}, \citenamefont {Bicket}, \citenamefont {Rembold},\ and\ \citenamefont {Haslinger}}]{preimesberger2025experimental}%
  \BibitemOpen
  \bibfield  {author} {\bibinfo {author} {\bibfnamefont {A.}~\bibnamefont {Preimesberger}}, \bibinfo {author} {\bibfnamefont {S.}~\bibnamefont {Bogdanov}}, \bibinfo {author} {\bibfnamefont {I.~C.}\ \bibnamefont {Bicket}}, \bibinfo {author} {\bibfnamefont {P.}~\bibnamefont {Rembold}},\ and\ \bibinfo {author} {\bibfnamefont {P.}~\bibnamefont {Haslinger}},\ }\bibfield  {journal} {\bibinfo  {journal} {arXiv preprint arXiv:2504.13163}\ }\href {https://doi.org/10.48550/arXiv.2504.13163} {10.48550/arXiv.2504.13163} (\bibinfo {year} {2025})\BibitemShut {NoStop}%
\bibitem [{\citenamefont {Kumar}\ \emph {et~al.}(2024)\citenamefont {Kumar}, \citenamefont {Lim}, \citenamefont {Rivera}, \citenamefont {Wong}, \citenamefont {Ang}, \citenamefont {Ang},\ and\ \citenamefont {Wong}}]{kumar2024strongly}%
  \BibitemOpen
  \bibfield  {author} {\bibinfo {author} {\bibfnamefont {S.}~\bibnamefont {Kumar}}, \bibinfo {author} {\bibfnamefont {J.}~\bibnamefont {Lim}}, \bibinfo {author} {\bibfnamefont {N.}~\bibnamefont {Rivera}}, \bibinfo {author} {\bibfnamefont {W.}~\bibnamefont {Wong}}, \bibinfo {author} {\bibfnamefont {Y.~S.}\ \bibnamefont {Ang}}, \bibinfo {author} {\bibfnamefont {L.~K.}\ \bibnamefont {Ang}},\ and\ \bibinfo {author} {\bibfnamefont {L.~J.}\ \bibnamefont {Wong}},\ }\href {https://doi.org/10.1126/sciadv.adm9563} {\bibfield  {journal} {\bibinfo  {journal} {Science Advances}\ }\textbf {\bibinfo {volume} {10}},\ \bibinfo {pages} {eadm9563} (\bibinfo {year} {2024})}\BibitemShut {NoStop}%
\bibitem [{\citenamefont {Baranes}\ \emph {et~al.}(2022)\citenamefont {Baranes}, \citenamefont {Ruimy}, \citenamefont {Gorlach},\ and\ \citenamefont {Kaminer}}]{baranes2022free}%
  \BibitemOpen
  \bibfield  {author} {\bibinfo {author} {\bibfnamefont {G.}~\bibnamefont {Baranes}}, \bibinfo {author} {\bibfnamefont {R.}~\bibnamefont {Ruimy}}, \bibinfo {author} {\bibfnamefont {A.}~\bibnamefont {Gorlach}},\ and\ \bibinfo {author} {\bibfnamefont {I.}~\bibnamefont {Kaminer}},\ }\href {https://doi.org/10.1038/s41534-022-00540-4} {\bibfield  {journal} {\bibinfo  {journal} {npj Quantum Information}\ }\textbf {\bibinfo {volume} {8}},\ \bibinfo {pages} {32} (\bibinfo {year} {2022})}\BibitemShut {NoStop}%
\bibitem [{\citenamefont {Piazza}\ \emph {et~al.}(2015)\citenamefont {Piazza}, \citenamefont {Lummen}, \citenamefont {Quinonez}, \citenamefont {Murooka}, \citenamefont {Reed}, \citenamefont {Barwick},\ and\ \citenamefont {Carbone}}]{piazza2015simultaneous}%
  \BibitemOpen
  \bibfield  {author} {\bibinfo {author} {\bibfnamefont {L.}~\bibnamefont {Piazza}}, \bibinfo {author} {\bibfnamefont {T.~T.}\ \bibnamefont {Lummen}}, \bibinfo {author} {\bibfnamefont {E.}~\bibnamefont {Quinonez}}, \bibinfo {author} {\bibfnamefont {Y.}~\bibnamefont {Murooka}}, \bibinfo {author} {\bibfnamefont {B.~W.}\ \bibnamefont {Reed}}, \bibinfo {author} {\bibfnamefont {B.}~\bibnamefont {Barwick}},\ and\ \bibinfo {author} {\bibfnamefont {F.}~\bibnamefont {Carbone}},\ }\href {https://doi.org/10.1038/ncomms7407} {\bibfield  {journal} {\bibinfo  {journal} {Nature communications}\ }\textbf {\bibinfo {volume} {6}},\ \bibinfo {pages} {6407} (\bibinfo {year} {2015})}\BibitemShut {NoStop}%
\bibitem [{\citenamefont {Wang}\ \emph {et~al.}(2020{\natexlab{b}})\citenamefont {Wang}, \citenamefont {Dahan}, \citenamefont {Shentcis}, \citenamefont {Kauffmann}, \citenamefont {Ben~Hayun}, \citenamefont {Reinhardt}, \citenamefont {Tsesses},\ and\ \citenamefont {Kaminer}}]{wang2020coherent}%
  \BibitemOpen
  \bibfield  {author} {\bibinfo {author} {\bibfnamefont {K.}~\bibnamefont {Wang}}, \bibinfo {author} {\bibfnamefont {R.}~\bibnamefont {Dahan}}, \bibinfo {author} {\bibfnamefont {M.}~\bibnamefont {Shentcis}}, \bibinfo {author} {\bibfnamefont {Y.}~\bibnamefont {Kauffmann}}, \bibinfo {author} {\bibfnamefont {A.}~\bibnamefont {Ben~Hayun}}, \bibinfo {author} {\bibfnamefont {O.}~\bibnamefont {Reinhardt}}, \bibinfo {author} {\bibfnamefont {S.}~\bibnamefont {Tsesses}},\ and\ \bibinfo {author} {\bibfnamefont {I.}~\bibnamefont {Kaminer}},\ }\href {https://doi.org/10.1038/s41586-020-2321-x} {\bibfield  {journal} {\bibinfo  {journal} {Nature}\ }\textbf {\bibinfo {volume} {582}},\ \bibinfo {pages} {50} (\bibinfo {year} {2020}{\natexlab{b}})}\BibitemShut {NoStop}%
\bibitem [{\citenamefont {Kfir}\ \emph {et~al.}(2020)\citenamefont {Kfir}, \citenamefont {Louren{\c{c}}o-Martins}, \citenamefont {Storeck}, \citenamefont {Sivis}, \citenamefont {Harvey}, \citenamefont {Kippenberg}, \citenamefont {Feist},\ and\ \citenamefont {Ropers}}]{kfir2020controlling}%
  \BibitemOpen
  \bibfield  {author} {\bibinfo {author} {\bibfnamefont {O.}~\bibnamefont {Kfir}}, \bibinfo {author} {\bibfnamefont {H.}~\bibnamefont {Louren{\c{c}}o-Martins}}, \bibinfo {author} {\bibfnamefont {G.}~\bibnamefont {Storeck}}, \bibinfo {author} {\bibfnamefont {M.}~\bibnamefont {Sivis}}, \bibinfo {author} {\bibfnamefont {T.~R.}\ \bibnamefont {Harvey}}, \bibinfo {author} {\bibfnamefont {T.~J.}\ \bibnamefont {Kippenberg}}, \bibinfo {author} {\bibfnamefont {A.}~\bibnamefont {Feist}},\ and\ \bibinfo {author} {\bibfnamefont {C.}~\bibnamefont {Ropers}},\ }\href {https://doi.org/10.1038/s41586-020-2320-y} {\bibfield  {journal} {\bibinfo  {journal} {Nature}\ }\textbf {\bibinfo {volume} {582}},\ \bibinfo {pages} {46} (\bibinfo {year} {2020})}\BibitemShut {NoStop}%
\bibitem [{\citenamefont {Henke}\ \emph {et~al.}(2021)\citenamefont {Henke}, \citenamefont {Raja}, \citenamefont {Feist}, \citenamefont {Huang}, \citenamefont {Arend}, \citenamefont {Yang}, \citenamefont {Kappert}, \citenamefont {Wang}, \citenamefont {M{\"o}ller}, \citenamefont {Pan} \emph {et~al.}}]{henke2021integrated}%
  \BibitemOpen
  \bibfield  {author} {\bibinfo {author} {\bibfnamefont {J.-W.}\ \bibnamefont {Henke}}, \bibinfo {author} {\bibfnamefont {A.~S.}\ \bibnamefont {Raja}}, \bibinfo {author} {\bibfnamefont {A.}~\bibnamefont {Feist}}, \bibinfo {author} {\bibfnamefont {G.}~\bibnamefont {Huang}}, \bibinfo {author} {\bibfnamefont {G.}~\bibnamefont {Arend}}, \bibinfo {author} {\bibfnamefont {Y.}~\bibnamefont {Yang}}, \bibinfo {author} {\bibfnamefont {F.~J.}\ \bibnamefont {Kappert}}, \bibinfo {author} {\bibfnamefont {R.~N.}\ \bibnamefont {Wang}}, \bibinfo {author} {\bibfnamefont {M.}~\bibnamefont {M{\"o}ller}}, \bibinfo {author} {\bibfnamefont {J.}~\bibnamefont {Pan}}, \emph {et~al.},\ }\href {https://doi.org/10.1038/s41586-021-04197-5} {\bibfield  {journal} {\bibinfo  {journal} {Nature}\ }\textbf {\bibinfo {volume} {600}},\ \bibinfo {pages} {653} (\bibinfo {year} {2021})}\BibitemShut {NoStop}%
\bibitem [{\citenamefont {Kurman}\ \emph {et~al.}(2021)\citenamefont {Kurman}, \citenamefont {Dahan}, \citenamefont {Sheinfux}, \citenamefont {Wang}, \citenamefont {Yannai}, \citenamefont {Adiv}, \citenamefont {Reinhardt}, \citenamefont {Tizei}, \citenamefont {Woo}, \citenamefont {Li} \emph {et~al.}}]{kurman2021spatiotemporal}%
  \BibitemOpen
  \bibfield  {author} {\bibinfo {author} {\bibfnamefont {Y.}~\bibnamefont {Kurman}}, \bibinfo {author} {\bibfnamefont {R.}~\bibnamefont {Dahan}}, \bibinfo {author} {\bibfnamefont {H.~H.}\ \bibnamefont {Sheinfux}}, \bibinfo {author} {\bibfnamefont {K.}~\bibnamefont {Wang}}, \bibinfo {author} {\bibfnamefont {M.}~\bibnamefont {Yannai}}, \bibinfo {author} {\bibfnamefont {Y.}~\bibnamefont {Adiv}}, \bibinfo {author} {\bibfnamefont {O.}~\bibnamefont {Reinhardt}}, \bibinfo {author} {\bibfnamefont {L.~H.}\ \bibnamefont {Tizei}}, \bibinfo {author} {\bibfnamefont {S.~Y.}\ \bibnamefont {Woo}}, \bibinfo {author} {\bibfnamefont {J.}~\bibnamefont {Li}}, \emph {et~al.},\ }\href {https://doi.org/10.1126/science.abg9015} {\bibfield  {journal} {\bibinfo  {journal} {Science}\ }\textbf {\bibinfo {volume} {372}},\ \bibinfo {pages} {1181} (\bibinfo {year} {2021})}\BibitemShut {NoStop}%
\bibitem [{\citenamefont {Yang}\ \emph {et~al.}(2024)\citenamefont {Yang}, \citenamefont {Henke}, \citenamefont {Raja}, \citenamefont {Kappert}, \citenamefont {Huang}, \citenamefont {Arend}, \citenamefont {Qiu}, \citenamefont {Feist}, \citenamefont {Wang}, \citenamefont {Tusnin} \emph {et~al.}}]{yang2024free}%
  \BibitemOpen
  \bibfield  {author} {\bibinfo {author} {\bibfnamefont {Y.}~\bibnamefont {Yang}}, \bibinfo {author} {\bibfnamefont {J.-W.}\ \bibnamefont {Henke}}, \bibinfo {author} {\bibfnamefont {A.~S.}\ \bibnamefont {Raja}}, \bibinfo {author} {\bibfnamefont {F.~J.}\ \bibnamefont {Kappert}}, \bibinfo {author} {\bibfnamefont {G.}~\bibnamefont {Huang}}, \bibinfo {author} {\bibfnamefont {G.}~\bibnamefont {Arend}}, \bibinfo {author} {\bibfnamefont {Z.}~\bibnamefont {Qiu}}, \bibinfo {author} {\bibfnamefont {A.}~\bibnamefont {Feist}}, \bibinfo {author} {\bibfnamefont {R.~N.}\ \bibnamefont {Wang}}, \bibinfo {author} {\bibfnamefont {A.}~\bibnamefont {Tusnin}}, \emph {et~al.},\ }\href {https://doi.org/10.1126/science.adk2489} {\bibfield  {journal} {\bibinfo  {journal} {Science}\ }\textbf {\bibinfo {volume} {383}},\ \bibinfo {pages} {168} (\bibinfo {year} {2024})}\BibitemShut {NoStop}%
\bibitem [{\citenamefont {Feist}\ \emph {et~al.}(2015)\citenamefont {Feist}, \citenamefont {Echternkamp}, \citenamefont {Schauss}, \citenamefont {Yalunin}, \citenamefont {Sch{\"a}fer},\ and\ \citenamefont {Ropers}}]{feist2015quantum}%
  \BibitemOpen
  \bibfield  {author} {\bibinfo {author} {\bibfnamefont {A.}~\bibnamefont {Feist}}, \bibinfo {author} {\bibfnamefont {K.~E.}\ \bibnamefont {Echternkamp}}, \bibinfo {author} {\bibfnamefont {J.}~\bibnamefont {Schauss}}, \bibinfo {author} {\bibfnamefont {S.~V.}\ \bibnamefont {Yalunin}}, \bibinfo {author} {\bibfnamefont {S.}~\bibnamefont {Sch{\"a}fer}},\ and\ \bibinfo {author} {\bibfnamefont {C.}~\bibnamefont {Ropers}},\ }\href {https://doi.org/10.1038/nature14463} {\bibfield  {journal} {\bibinfo  {journal} {Nature}\ }\textbf {\bibinfo {volume} {521}},\ \bibinfo {pages} {200} (\bibinfo {year} {2015})}\BibitemShut {NoStop}%
\bibitem [{\citenamefont {Morimoto}\ and\ \citenamefont {Baum}(2018)}]{morimoto2018diffraction}%
  \BibitemOpen
  \bibfield  {author} {\bibinfo {author} {\bibfnamefont {Y.}~\bibnamefont {Morimoto}}\ and\ \bibinfo {author} {\bibfnamefont {P.}~\bibnamefont {Baum}},\ }\href {https://doi.org/10.1038/s41567-017-0007-6} {\bibfield  {journal} {\bibinfo  {journal} {Nature Physics}\ }\textbf {\bibinfo {volume} {14}},\ \bibinfo {pages} {252} (\bibinfo {year} {2018})}\BibitemShut {NoStop}%
\bibitem [{\citenamefont {Pan}\ \emph {et~al.}(2019)\citenamefont {Pan}, \citenamefont {Zhang},\ and\ \citenamefont {Gover}}]{pan2019anomalous}%
  \BibitemOpen
  \bibfield  {author} {\bibinfo {author} {\bibfnamefont {Y.}~\bibnamefont {Pan}}, \bibinfo {author} {\bibfnamefont {B.}~\bibnamefont {Zhang}},\ and\ \bibinfo {author} {\bibfnamefont {A.}~\bibnamefont {Gover}},\ }\href {https://doi.org/10.1103/PhysRevLett.122.183204} {\bibfield  {journal} {\bibinfo  {journal} {Physical Review Letters}\ }\textbf {\bibinfo {volume} {122}},\ \bibinfo {pages} {183204} (\bibinfo {year} {2019})}\BibitemShut {NoStop}%
\bibitem [{\citenamefont {Vanacore}\ \emph {et~al.}(2019)\citenamefont {Vanacore}, \citenamefont {Berruto}, \citenamefont {Madan}, \citenamefont {Pomarico}, \citenamefont {Biagioni}, \citenamefont {Lamb}, \citenamefont {McGrouther}, \citenamefont {Reinhardt}, \citenamefont {Kaminer}, \citenamefont {Barwick} \emph {et~al.}}]{vanacore2019ultrafast}%
  \BibitemOpen
  \bibfield  {author} {\bibinfo {author} {\bibfnamefont {G.~M.}\ \bibnamefont {Vanacore}}, \bibinfo {author} {\bibfnamefont {G.}~\bibnamefont {Berruto}}, \bibinfo {author} {\bibfnamefont {I.}~\bibnamefont {Madan}}, \bibinfo {author} {\bibfnamefont {E.}~\bibnamefont {Pomarico}}, \bibinfo {author} {\bibfnamefont {P.}~\bibnamefont {Biagioni}}, \bibinfo {author} {\bibfnamefont {R.}~\bibnamefont {Lamb}}, \bibinfo {author} {\bibfnamefont {D.}~\bibnamefont {McGrouther}}, \bibinfo {author} {\bibfnamefont {O.}~\bibnamefont {Reinhardt}}, \bibinfo {author} {\bibfnamefont {I.}~\bibnamefont {Kaminer}}, \bibinfo {author} {\bibfnamefont {B.}~\bibnamefont {Barwick}}, \emph {et~al.},\ }\href {https://doi.org/10.1038/s41563-019-0336-1} {\bibfield  {journal} {\bibinfo  {journal} {Nature materials}\ }\textbf {\bibinfo {volume} {18}},\ \bibinfo {pages} {573} (\bibinfo {year} {2019})}\BibitemShut {NoStop}%
\bibitem [{\citenamefont {Dahan}\ \emph {et~al.}(2021)\citenamefont {Dahan}, \citenamefont {Gorlach}, \citenamefont {Haeusler}, \citenamefont {Karnieli}, \citenamefont {Eyal}, \citenamefont {Yousefi}, \citenamefont {Segev}, \citenamefont {Arie}, \citenamefont {Eisenstein}, \citenamefont {Hommelhoff} \emph {et~al.}}]{dahan2021imprinting}%
  \BibitemOpen
  \bibfield  {author} {\bibinfo {author} {\bibfnamefont {R.}~\bibnamefont {Dahan}}, \bibinfo {author} {\bibfnamefont {A.}~\bibnamefont {Gorlach}}, \bibinfo {author} {\bibfnamefont {U.}~\bibnamefont {Haeusler}}, \bibinfo {author} {\bibfnamefont {A.}~\bibnamefont {Karnieli}}, \bibinfo {author} {\bibfnamefont {O.}~\bibnamefont {Eyal}}, \bibinfo {author} {\bibfnamefont {P.}~\bibnamefont {Yousefi}}, \bibinfo {author} {\bibfnamefont {M.}~\bibnamefont {Segev}}, \bibinfo {author} {\bibfnamefont {A.}~\bibnamefont {Arie}}, \bibinfo {author} {\bibfnamefont {G.}~\bibnamefont {Eisenstein}}, \bibinfo {author} {\bibfnamefont {P.}~\bibnamefont {Hommelhoff}}, \emph {et~al.},\ }\href {https://doi.org/10.1126/science.abj7128} {\bibfield  {journal} {\bibinfo  {journal} {Science}\ }\textbf {\bibinfo {volume} {373}},\ \bibinfo {pages} {eabj7128} (\bibinfo {year} {2021})}\BibitemShut {NoStop}%
\bibitem [{\citenamefont {Nabben}\ \emph {et~al.}(2023)\citenamefont {Nabben}, \citenamefont {Kuttruff}, \citenamefont {Stolz}, \citenamefont {Ryabov},\ and\ \citenamefont {Baum}}]{nabben2023attosecond}%
  \BibitemOpen
  \bibfield  {author} {\bibinfo {author} {\bibfnamefont {D.}~\bibnamefont {Nabben}}, \bibinfo {author} {\bibfnamefont {J.}~\bibnamefont {Kuttruff}}, \bibinfo {author} {\bibfnamefont {L.}~\bibnamefont {Stolz}}, \bibinfo {author} {\bibfnamefont {A.}~\bibnamefont {Ryabov}},\ and\ \bibinfo {author} {\bibfnamefont {P.}~\bibnamefont {Baum}},\ }\href {https://doi.org/10.1038/s41586-023-06074-9} {\bibfield  {journal} {\bibinfo  {journal} {Nature}\ }\textbf {\bibinfo {volume} {619}},\ \bibinfo {pages} {63} (\bibinfo {year} {2023})}\BibitemShut {NoStop}%
\bibitem [{\citenamefont {Gaida}\ \emph {et~al.}(2024)\citenamefont {Gaida}, \citenamefont {Louren{\c{c}}o-Martins}, \citenamefont {Sivis}, \citenamefont {Rittmann}, \citenamefont {Feist}, \citenamefont {Garc{\'\i}a~de Abajo},\ and\ \citenamefont {Ropers}}]{gaida2024attosecond}%
  \BibitemOpen
  \bibfield  {author} {\bibinfo {author} {\bibfnamefont {J.~H.}\ \bibnamefont {Gaida}}, \bibinfo {author} {\bibfnamefont {H.}~\bibnamefont {Louren{\c{c}}o-Martins}}, \bibinfo {author} {\bibfnamefont {M.}~\bibnamefont {Sivis}}, \bibinfo {author} {\bibfnamefont {T.}~\bibnamefont {Rittmann}}, \bibinfo {author} {\bibfnamefont {A.}~\bibnamefont {Feist}}, \bibinfo {author} {\bibfnamefont {F.~J.}\ \bibnamefont {Garc{\'\i}a~de Abajo}},\ and\ \bibinfo {author} {\bibfnamefont {C.}~\bibnamefont {Ropers}},\ }\href {https://doi.org/10.1038/s41566-024-01380-8} {\bibfield  {journal} {\bibinfo  {journal} {Nature Photonics}\ }\textbf {\bibinfo {volume} {18}},\ \bibinfo {pages} {509} (\bibinfo {year} {2024})}\BibitemShut {NoStop}%
\bibitem [{\citenamefont {Fang}\ \emph {et~al.}(2024)\citenamefont {Fang}, \citenamefont {Kuttruff}, \citenamefont {Nabben},\ and\ \citenamefont {Baum}}]{fang2024structured}%
  \BibitemOpen
  \bibfield  {author} {\bibinfo {author} {\bibfnamefont {Y.}~\bibnamefont {Fang}}, \bibinfo {author} {\bibfnamefont {J.}~\bibnamefont {Kuttruff}}, \bibinfo {author} {\bibfnamefont {D.}~\bibnamefont {Nabben}},\ and\ \bibinfo {author} {\bibfnamefont {P.}~\bibnamefont {Baum}},\ }\href {https://doi.org/10.1126/science.adp9143} {\bibfield  {journal} {\bibinfo  {journal} {Science}\ }\textbf {\bibinfo {volume} {385}},\ \bibinfo {pages} {183} (\bibinfo {year} {2024})}\BibitemShut {NoStop}%
\bibitem [{\citenamefont {Bendana}\ \emph {et~al.}(2011)\citenamefont {Bendana}, \citenamefont {Polman},\ and\ \citenamefont {Garc{\'\i}a~de Abajo}}]{bendana2011single}%
  \BibitemOpen
  \bibfield  {author} {\bibinfo {author} {\bibfnamefont {X.}~\bibnamefont {Bendana}}, \bibinfo {author} {\bibfnamefont {A.}~\bibnamefont {Polman}},\ and\ \bibinfo {author} {\bibfnamefont {F.~J.}\ \bibnamefont {Garc{\'\i}a~de Abajo}},\ }\href {https://doi.org/10.1021/nl1034732} {\bibfield  {journal} {\bibinfo  {journal} {Nano letters}\ }\textbf {\bibinfo {volume} {11}},\ \bibinfo {pages} {5099} (\bibinfo {year} {2011})}\BibitemShut {NoStop}%
\bibitem [{\citenamefont {Feist}\ \emph {et~al.}(2022)\citenamefont {Feist}, \citenamefont {Huang}, \citenamefont {Arend}, \citenamefont {Yang}, \citenamefont {Henke}, \citenamefont {Raja}, \citenamefont {Kappert}, \citenamefont {Wang}, \citenamefont {Louren{\c{c}}o-Martins}, \citenamefont {Qiu} \emph {et~al.}}]{feist2022cavity}%
  \BibitemOpen
  \bibfield  {author} {\bibinfo {author} {\bibfnamefont {A.}~\bibnamefont {Feist}}, \bibinfo {author} {\bibfnamefont {G.}~\bibnamefont {Huang}}, \bibinfo {author} {\bibfnamefont {G.}~\bibnamefont {Arend}}, \bibinfo {author} {\bibfnamefont {Y.}~\bibnamefont {Yang}}, \bibinfo {author} {\bibfnamefont {J.-W.}\ \bibnamefont {Henke}}, \bibinfo {author} {\bibfnamefont {A.~S.}\ \bibnamefont {Raja}}, \bibinfo {author} {\bibfnamefont {F.~J.}\ \bibnamefont {Kappert}}, \bibinfo {author} {\bibfnamefont {R.~N.}\ \bibnamefont {Wang}}, \bibinfo {author} {\bibfnamefont {H.}~\bibnamefont {Louren{\c{c}}o-Martins}}, \bibinfo {author} {\bibfnamefont {Z.}~\bibnamefont {Qiu}}, \emph {et~al.},\ }\href {https://doi.org/10.1126/sciadv.abq4947} {\bibfield  {journal} {\bibinfo  {journal} {Science}\ }\textbf {\bibinfo {volume} {377}},\ \bibinfo {pages} {777} (\bibinfo {year} {2022})}\BibitemShut {NoStop}%
\bibitem [{\citenamefont {Dahan}\ \emph {et~al.}(2023)\citenamefont {Dahan}, \citenamefont {Baranes}, \citenamefont {Gorlach}, \citenamefont {Ruimy}, \citenamefont {Rivera},\ and\ \citenamefont {Kaminer}}]{dahan2023creation}%
  \BibitemOpen
  \bibfield  {author} {\bibinfo {author} {\bibfnamefont {R.}~\bibnamefont {Dahan}}, \bibinfo {author} {\bibfnamefont {G.}~\bibnamefont {Baranes}}, \bibinfo {author} {\bibfnamefont {A.}~\bibnamefont {Gorlach}}, \bibinfo {author} {\bibfnamefont {R.}~\bibnamefont {Ruimy}}, \bibinfo {author} {\bibfnamefont {N.}~\bibnamefont {Rivera}},\ and\ \bibinfo {author} {\bibfnamefont {I.}~\bibnamefont {Kaminer}},\ }\href {https://doi.org/10.1103/PhysRevX.13.031001} {\bibfield  {journal} {\bibinfo  {journal} {Physical Review X}\ }\textbf {\bibinfo {volume} {13}},\ \bibinfo {pages} {031001} (\bibinfo {year} {2023})}\BibitemShut {NoStop}%
\bibitem [{\citenamefont {Arend}\ \emph {et~al.}(2025)\citenamefont {Arend}, \citenamefont {Huang}, \citenamefont {Feist}, \citenamefont {Yang}, \citenamefont {Henke}, \citenamefont {Qiu}, \citenamefont {Jeng}, \citenamefont {Raja}, \citenamefont {Haindl}, \citenamefont {Wang} \emph {et~al.}}]{arend2024electrons}%
  \BibitemOpen
  \bibfield  {author} {\bibinfo {author} {\bibfnamefont {G.}~\bibnamefont {Arend}}, \bibinfo {author} {\bibfnamefont {G.}~\bibnamefont {Huang}}, \bibinfo {author} {\bibfnamefont {A.}~\bibnamefont {Feist}}, \bibinfo {author} {\bibfnamefont {Y.}~\bibnamefont {Yang}}, \bibinfo {author} {\bibfnamefont {J.-W.}\ \bibnamefont {Henke}}, \bibinfo {author} {\bibfnamefont {Z.}~\bibnamefont {Qiu}}, \bibinfo {author} {\bibfnamefont {H.}~\bibnamefont {Jeng}}, \bibinfo {author} {\bibfnamefont {A.~S.}\ \bibnamefont {Raja}}, \bibinfo {author} {\bibfnamefont {R.}~\bibnamefont {Haindl}}, \bibinfo {author} {\bibfnamefont {R.~N.}\ \bibnamefont {Wang}}, \emph {et~al.},\ }\href {https://doi.org/10.1038/s41567-025-03033-1} {\bibfield  {journal} {\bibinfo  {journal} {Nature Physics}\ ,\ \bibinfo {pages} {1}} (\bibinfo {year} {2025})}\BibitemShut {NoStop}%
\bibitem [{\citenamefont {Batelaan}(2007)}]{batelaan2007colloquium}%
  \BibitemOpen
  \bibfield  {author} {\bibinfo {author} {\bibfnamefont {H.}~\bibnamefont {Batelaan}},\ }\href {https://doi.org/10.1103/RevModPhys.79.929} {\bibfield  {journal} {\bibinfo  {journal} {Reviews of Modern Physics}\ }\textbf {\bibinfo {volume} {79}},\ \bibinfo {pages} {929} (\bibinfo {year} {2007})}\BibitemShut {NoStop}%
\bibitem [{\citenamefont {Koz{\'a}k}\ \emph {et~al.}(2018{\natexlab{a}})\citenamefont {Koz{\'a}k}, \citenamefont {Eckstein}, \citenamefont {Sch{\"o}nenberger},\ and\ \citenamefont {Hommelhoff}}]{kozak2018inelastic}%
  \BibitemOpen
  \bibfield  {author} {\bibinfo {author} {\bibfnamefont {M.}~\bibnamefont {Koz{\'a}k}}, \bibinfo {author} {\bibfnamefont {T.}~\bibnamefont {Eckstein}}, \bibinfo {author} {\bibfnamefont {N.}~\bibnamefont {Sch{\"o}nenberger}},\ and\ \bibinfo {author} {\bibfnamefont {P.}~\bibnamefont {Hommelhoff}},\ }\href {https://doi.org/10.1038/nphys4282} {\bibfield  {journal} {\bibinfo  {journal} {Nature Physics}\ }\textbf {\bibinfo {volume} {14}},\ \bibinfo {pages} {121} (\bibinfo {year} {2018}{\natexlab{a}})}\BibitemShut {NoStop}%
\bibitem [{\citenamefont {Koz{\'a}k}\ \emph {et~al.}(2018{\natexlab{b}})\citenamefont {Koz{\'a}k}, \citenamefont {Sch{\"o}nenberger},\ and\ \citenamefont {Hommelhoff}}]{kozak2018ponderomotive}%
  \BibitemOpen
  \bibfield  {author} {\bibinfo {author} {\bibfnamefont {M.}~\bibnamefont {Koz{\'a}k}}, \bibinfo {author} {\bibfnamefont {N.}~\bibnamefont {Sch{\"o}nenberger}},\ and\ \bibinfo {author} {\bibfnamefont {P.}~\bibnamefont {Hommelhoff}},\ }\href {https://doi.org/10.1103/PhysRevLett.120.103203} {\bibfield  {journal} {\bibinfo  {journal} {Physical review letters}\ }\textbf {\bibinfo {volume} {120}},\ \bibinfo {pages} {103203} (\bibinfo {year} {2018}{\natexlab{b}})}\BibitemShut {NoStop}%
\bibitem [{\citenamefont {Garc{\'\i}a~de Abajo}\ and\ \citenamefont {Kone{\v{c}}n{\'a}}(2021)}]{garcia2021opticalprl}%
  \BibitemOpen
  \bibfield  {author} {\bibinfo {author} {\bibfnamefont {F.~J.}\ \bibnamefont {Garc{\'\i}a~de Abajo}}\ and\ \bibinfo {author} {\bibfnamefont {A.}~\bibnamefont {Kone{\v{c}}n{\'a}}},\ }\href@noop {} {\bibfield  {journal} {\bibinfo  {journal} {Physical Review Letters}\ }\textbf {\bibinfo {volume} {126}},\ \bibinfo {pages} {123901} (\bibinfo {year} {2021})}\BibitemShut {NoStop}%
\bibitem [{\citenamefont {Di~Giulio}\ and\ \citenamefont {Garc{\'\i}a~de Abajo}(2022)}]{di2022optical}%
  \BibitemOpen
  \bibfield  {author} {\bibinfo {author} {\bibfnamefont {V.}~\bibnamefont {Di~Giulio}}\ and\ \bibinfo {author} {\bibfnamefont {F.~J.}\ \bibnamefont {Garc{\'\i}a~de Abajo}},\ }\href {https://doi.org/10.1515/nanoph-2022-0481} {\bibfield  {journal} {\bibinfo  {journal} {Nanophotonics}\ }\textbf {\bibinfo {volume} {11}},\ \bibinfo {pages} {4659} (\bibinfo {year} {2022})}\BibitemShut {NoStop}%
\bibitem [{\citenamefont {Tsarev}\ \emph {et~al.}(2023)\citenamefont {Tsarev}, \citenamefont {Thurner},\ and\ \citenamefont {Baum}}]{tsarev2023nonlinear}%
  \BibitemOpen
  \bibfield  {author} {\bibinfo {author} {\bibfnamefont {M.}~\bibnamefont {Tsarev}}, \bibinfo {author} {\bibfnamefont {J.~W.}\ \bibnamefont {Thurner}},\ and\ \bibinfo {author} {\bibfnamefont {P.}~\bibnamefont {Baum}},\ }\href {https://doi.org/10.1038/s41567-023-02092-6} {\bibfield  {journal} {\bibinfo  {journal} {Nature Physics}\ }\textbf {\bibinfo {volume} {19}},\ \bibinfo {pages} {1350} (\bibinfo {year} {2023})}\BibitemShut {NoStop}%
\bibitem [{\citenamefont {Talebi}(2020)}]{talebi2020strong}%
  \BibitemOpen
  \bibfield  {author} {\bibinfo {author} {\bibfnamefont {N.}~\bibnamefont {Talebi}},\ }\href {https://doi.org/10.1103/PhysRevLett.125.080401} {\bibfield  {journal} {\bibinfo  {journal} {Physical Review Letters}\ }\textbf {\bibinfo {volume} {125}},\ \bibinfo {pages} {080401} (\bibinfo {year} {2020})}\BibitemShut {NoStop}%
\bibitem [{\citenamefont {Garc{\'\i}a~de Abajo}\ \emph {et~al.}(2022)\citenamefont {Garc{\'\i}a~de Abajo}, \citenamefont {Dias},\ and\ \citenamefont {Di~Giulio}}]{garcia2022complete}%
  \BibitemOpen
  \bibfield  {author} {\bibinfo {author} {\bibfnamefont {F.~J.}\ \bibnamefont {Garc{\'\i}a~de Abajo}}, \bibinfo {author} {\bibfnamefont {E.~J.}\ \bibnamefont {Dias}},\ and\ \bibinfo {author} {\bibfnamefont {V.}~\bibnamefont {Di~Giulio}},\ }\href {https://doi.org/10.1103/PhysRevLett.129.093401} {\bibfield  {journal} {\bibinfo  {journal} {Physical Review Letters}\ }\textbf {\bibinfo {volume} {129}},\ \bibinfo {pages} {093401} (\bibinfo {year} {2022})}\BibitemShut {NoStop}%
\bibitem [{\citenamefont {Karnieli}\ and\ \citenamefont {Fan}(2023)}]{karnieli2023jaynes}%
  \BibitemOpen
  \bibfield  {author} {\bibinfo {author} {\bibfnamefont {A.}~\bibnamefont {Karnieli}}\ and\ \bibinfo {author} {\bibfnamefont {S.}~\bibnamefont {Fan}},\ }\href {https://doi.org/10.1126/sciadv.adh2425} {\bibfield  {journal} {\bibinfo  {journal} {Science advances}\ }\textbf {\bibinfo {volume} {9}},\ \bibinfo {pages} {eadh2425} (\bibinfo {year} {2023})}\BibitemShut {NoStop}%
\bibitem [{\citenamefont {Eldar}\ \emph {et~al.}(2024)\citenamefont {Eldar}, \citenamefont {Chen}, \citenamefont {Pan},\ and\ \citenamefont {Kr{\"u}ger}}]{eldar2024self}%
  \BibitemOpen
  \bibfield  {author} {\bibinfo {author} {\bibfnamefont {M.}~\bibnamefont {Eldar}}, \bibinfo {author} {\bibfnamefont {Z.}~\bibnamefont {Chen}}, \bibinfo {author} {\bibfnamefont {Y.}~\bibnamefont {Pan}},\ and\ \bibinfo {author} {\bibfnamefont {M.}~\bibnamefont {Kr{\"u}ger}},\ }\href {https://doi.org/PhysRevLett.132.035001} {\bibfield  {journal} {\bibinfo  {journal} {Physical Review Letters}\ }\textbf {\bibinfo {volume} {132}},\ \bibinfo {pages} {035001} (\bibinfo {year} {2024})}\BibitemShut {NoStop}%
\bibitem [{\citenamefont {Karnieli}\ \emph {et~al.}(2024{\natexlab{a}})\citenamefont {Karnieli}, \citenamefont {Roques-Carmes}, \citenamefont {Rivera},\ and\ \citenamefont {Fan}}]{karnieli2024strong}%
  \BibitemOpen
  \bibfield  {author} {\bibinfo {author} {\bibfnamefont {A.}~\bibnamefont {Karnieli}}, \bibinfo {author} {\bibfnamefont {C.}~\bibnamefont {Roques-Carmes}}, \bibinfo {author} {\bibfnamefont {N.}~\bibnamefont {Rivera}},\ and\ \bibinfo {author} {\bibfnamefont {S.}~\bibnamefont {Fan}},\ }\href {https://doi.org/10.1021/acsphotonics.4c00908} {\bibfield  {journal} {\bibinfo  {journal} {ACS Photonics}\ }\textbf {\bibinfo {volume} {11}},\ \bibinfo {pages} {3401} (\bibinfo {year} {2024}{\natexlab{a}})}\BibitemShut {NoStop}%
\bibitem [{\citenamefont {Karnieli}\ \emph {et~al.}(2024{\natexlab{b}})\citenamefont {Karnieli}, \citenamefont {Tsesses}, \citenamefont {Yu}, \citenamefont {Rivera}, \citenamefont {Arie}, \citenamefont {Kaminer},\ and\ \citenamefont {Fan}}]{karnieli2024universal}%
  \BibitemOpen
  \bibfield  {author} {\bibinfo {author} {\bibfnamefont {A.}~\bibnamefont {Karnieli}}, \bibinfo {author} {\bibfnamefont {S.}~\bibnamefont {Tsesses}}, \bibinfo {author} {\bibfnamefont {R.}~\bibnamefont {Yu}}, \bibinfo {author} {\bibfnamefont {N.}~\bibnamefont {Rivera}}, \bibinfo {author} {\bibfnamefont {A.}~\bibnamefont {Arie}}, \bibinfo {author} {\bibfnamefont {I.}~\bibnamefont {Kaminer}},\ and\ \bibinfo {author} {\bibfnamefont {S.}~\bibnamefont {Fan}},\ }\href {https://doi.org/10.1103/PRXQuantum.5.010339} {\bibfield  {journal} {\bibinfo  {journal} {PRX Quantum}\ }\textbf {\bibinfo {volume} {5}},\ \bibinfo {pages} {010339} (\bibinfo {year} {2024}{\natexlab{b}})}\BibitemShut {NoStop}%
\bibitem [{\citenamefont {Sirotin}\ \emph {et~al.}(2024)\citenamefont {Sirotin}, \citenamefont {Rasputnyi}, \citenamefont {Chlouba}, \citenamefont {Shiloh},\ and\ \citenamefont {Hommelhoff}}]{sirotin2024quantum}%
  \BibitemOpen
  \bibfield  {author} {\bibinfo {author} {\bibfnamefont {M.}~\bibnamefont {Sirotin}}, \bibinfo {author} {\bibfnamefont {A.}~\bibnamefont {Rasputnyi}}, \bibinfo {author} {\bibfnamefont {T.}~\bibnamefont {Chlouba}}, \bibinfo {author} {\bibfnamefont {R.}~\bibnamefont {Shiloh}},\ and\ \bibinfo {author} {\bibfnamefont {P.}~\bibnamefont {Hommelhoff}},\ }\bibfield  {journal} {\bibinfo  {journal} {arXiv preprint arXiv:2405.06560}\ }\href {https://doi.org/10.48550/arXiv.2405.06560} {10.48550/arXiv.2405.06560} (\bibinfo {year} {2024})\BibitemShut {NoStop}%
\bibitem [{\citenamefont {Synanidis}\ \emph {et~al.}(2024)\citenamefont {Synanidis}, \citenamefont {Gon{\c{c}}alves}, \citenamefont {Ropers},\ and\ \citenamefont {de~Abajo}}]{synanidis2024quantum}%
  \BibitemOpen
  \bibfield  {author} {\bibinfo {author} {\bibfnamefont {A.~P.}\ \bibnamefont {Synanidis}}, \bibinfo {author} {\bibfnamefont {P.}~\bibnamefont {Gon{\c{c}}alves}}, \bibinfo {author} {\bibfnamefont {C.}~\bibnamefont {Ropers}},\ and\ \bibinfo {author} {\bibfnamefont {F.~J.~G.}\ \bibnamefont {de~Abajo}},\ }\href {https://doi.org/10.1126/sciadv.adp4096} {\bibfield  {journal} {\bibinfo  {journal} {Science Advances}\ }\textbf {\bibinfo {volume} {10}},\ \bibinfo {pages} {eadp4096} (\bibinfo {year} {2024})}\BibitemShut {NoStop}%
\bibitem [{\citenamefont {Konecna}\ \emph {et~al.}(2020)\citenamefont {Konecna}, \citenamefont {Di~Giulio}, \citenamefont {Mkhitaryan}, \citenamefont {Ropers},\ and\ \citenamefont {Garc{\'\i}a~de Abajo}}]{konecna2020nanoscale}%
  \BibitemOpen
  \bibfield  {author} {\bibinfo {author} {\bibfnamefont {A.}~\bibnamefont {Konecna}}, \bibinfo {author} {\bibfnamefont {V.}~\bibnamefont {Di~Giulio}}, \bibinfo {author} {\bibfnamefont {V.}~\bibnamefont {Mkhitaryan}}, \bibinfo {author} {\bibfnamefont {C.}~\bibnamefont {Ropers}},\ and\ \bibinfo {author} {\bibfnamefont {F.~J.}\ \bibnamefont {Garc{\'\i}a~de Abajo}},\ }\href {https://doi.org/10.1021/acsphotonics.0c00326} {\bibfield  {journal} {\bibinfo  {journal} {ACS Photonics}\ }\textbf {\bibinfo {volume} {7}},\ \bibinfo {pages} {1290} (\bibinfo {year} {2020})}\BibitemShut {NoStop}%
\bibitem [{\citenamefont {Prelat}\ \emph {et~al.}(2025)\citenamefont {Prelat}, \citenamefont {Dias},\ and\ \citenamefont {de~Abajo}}]{prelat2025wave}%
  \BibitemOpen
  \bibfield  {author} {\bibinfo {author} {\bibfnamefont {L.}~\bibnamefont {Prelat}}, \bibinfo {author} {\bibfnamefont {E.~J.}\ \bibnamefont {Dias}},\ and\ \bibinfo {author} {\bibfnamefont {F.}~\bibnamefont {de~Abajo}},\ }\bibfield  {journal} {\bibinfo  {journal} {arXiv preprint arXiv:2508.00560}\ }\href {https://doi.org/10.48550/arXiv.2508.00560} {10.48550/arXiv.2508.00560} (\bibinfo {year} {2025})\BibitemShut {NoStop}%
\bibitem [{\citenamefont {Cox}\ and\ \citenamefont {Garc{\'\i}a~Abajo}(2020)}]{cox2020nonlinear}%
  \BibitemOpen
  \bibfield  {author} {\bibinfo {author} {\bibfnamefont {J.~D.}\ \bibnamefont {Cox}}\ and\ \bibinfo {author} {\bibfnamefont {F.~J.}\ \bibnamefont {Garc{\'\i}a~Abajo}},\ }\href {https://doi.org/10.1021/acs.nanolett.0c00538} {\bibfield  {journal} {\bibinfo  {journal} {Nano Letters}\ }\textbf {\bibinfo {volume} {20}},\ \bibinfo {pages} {4792} (\bibinfo {year} {2020})}\BibitemShut {NoStop}%
\bibitem [{\citenamefont {Boyd}(2020)}]{boyd2008nonlinear}%
  \BibitemOpen
  \bibfield  {author} {\bibinfo {author} {\bibfnamefont {R.~W.}\ \bibnamefont {Boyd}},\ }\href {https://doi.org/10.1016/c2015-0-05510-1} {\emph {\bibinfo {title} {Nonlinear Optics}}},\ \bibinfo {edition} {4th}\ ed.\ (\bibinfo  {publisher} {Elsevier},\ \bibinfo {year} {2020})\BibitemShut {NoStop}%
\bibitem [{\citenamefont {Scully}\ and\ \citenamefont {Zubairy}(1997)}]{scully1997quantum}%
  \BibitemOpen
  \bibfield  {author} {\bibinfo {author} {\bibfnamefont {M.~O.}\ \bibnamefont {Scully}}\ and\ \bibinfo {author} {\bibfnamefont {M.~S.}\ \bibnamefont {Zubairy}},\ }\href {https://doi.org/10.1017/cbo9780511813993} {\emph {\bibinfo {title} {Quantum Optics}}}\ (\bibinfo  {publisher} {Cambridge University Press},\ \bibinfo {year} {1997})\BibitemShut {NoStop}%
\bibitem [{\citenamefont {Navarrete-Benlloch}(2022)}]{navarrete2022introduction}%
  \BibitemOpen
  \bibfield  {author} {\bibinfo {author} {\bibfnamefont {C.}~\bibnamefont {Navarrete-Benlloch}},\ }\bibfield  {journal} {\bibinfo  {journal} {arXiv preprint arXiv:2203.13206}\ }\href {https://doi.org/10.48550/arXiv.2203.13206} {10.48550/arXiv.2203.13206} (\bibinfo {year} {2022})\BibitemShut {NoStop}%
\bibitem [{\citenamefont {Xie}\ \emph {et~al.}(2025)\citenamefont {Xie}, \citenamefont {Chen}, \citenamefont {Li}, \citenamefont {Yan}, \citenamefont {Chen}, \citenamefont {Lin}, \citenamefont {Kaminer}, \citenamefont {Miller},\ and\ \citenamefont {Yang}}]{xie2025maximal}%
  \BibitemOpen
  \bibfield  {author} {\bibinfo {author} {\bibfnamefont {Z.}~\bibnamefont {Xie}}, \bibinfo {author} {\bibfnamefont {Z.}~\bibnamefont {Chen}}, \bibinfo {author} {\bibfnamefont {H.}~\bibnamefont {Li}}, \bibinfo {author} {\bibfnamefont {Q.}~\bibnamefont {Yan}}, \bibinfo {author} {\bibfnamefont {H.}~\bibnamefont {Chen}}, \bibinfo {author} {\bibfnamefont {X.}~\bibnamefont {Lin}}, \bibinfo {author} {\bibfnamefont {I.}~\bibnamefont {Kaminer}}, \bibinfo {author} {\bibfnamefont {O.~D.}\ \bibnamefont {Miller}},\ and\ \bibinfo {author} {\bibfnamefont {Y.}~\bibnamefont {Yang}},\ }\href {https://doi.org/10.1103/PhysRevLett.134.043803} {\bibfield  {journal} {\bibinfo  {journal} {Physical Review Letters}\ }\textbf {\bibinfo {volume} {134}},\ \bibinfo {pages} {043803} (\bibinfo {year} {2025})}\BibitemShut {NoStop}%
\bibitem [{\citenamefont {Zhao}(2025)}]{zhao2025upper}%
  \BibitemOpen
  \bibfield  {author} {\bibinfo {author} {\bibfnamefont {Z.}~\bibnamefont {Zhao}},\ }\href {https://doi.org/10.1103/PhysRevLett.134.043804} {\bibfield  {journal} {\bibinfo  {journal} {Physical Review Letters}\ }\textbf {\bibinfo {volume} {134}},\ \bibinfo {pages} {043804} (\bibinfo {year} {2025})}\BibitemShut {NoStop}%
\bibitem [{\citenamefont {Tomoda}\ \emph {et~al.}(2024)\citenamefont {Tomoda}, \citenamefont {Machinaga}, \citenamefont {Takase}, \citenamefont {Harada}, \citenamefont {Kashiwazaki}, \citenamefont {Umeki}, \citenamefont {Miki}, \citenamefont {China}, \citenamefont {Yabuno}, \citenamefont {Terai} \emph {et~al.}}]{tomoda2024boosting}%
  \BibitemOpen
  \bibfield  {author} {\bibinfo {author} {\bibfnamefont {H.}~\bibnamefont {Tomoda}}, \bibinfo {author} {\bibfnamefont {A.}~\bibnamefont {Machinaga}}, \bibinfo {author} {\bibfnamefont {K.}~\bibnamefont {Takase}}, \bibinfo {author} {\bibfnamefont {J.}~\bibnamefont {Harada}}, \bibinfo {author} {\bibfnamefont {T.}~\bibnamefont {Kashiwazaki}}, \bibinfo {author} {\bibfnamefont {T.}~\bibnamefont {Umeki}}, \bibinfo {author} {\bibfnamefont {S.}~\bibnamefont {Miki}}, \bibinfo {author} {\bibfnamefont {F.}~\bibnamefont {China}}, \bibinfo {author} {\bibfnamefont {M.}~\bibnamefont {Yabuno}}, \bibinfo {author} {\bibfnamefont {H.}~\bibnamefont {Terai}}, \emph {et~al.},\ }\href {https://doi.org/10.1103/PhysRevA.110.033717} {\bibfield  {journal} {\bibinfo  {journal} {Physical Review A}\ }\textbf {\bibinfo {volume} {110}},\ \bibinfo {pages} {033717} (\bibinfo {year} {2024})}\BibitemShut {NoStop}%
\bibitem [{\citenamefont {Korolev}\ \emph {et~al.}(2024)\citenamefont {Korolev}, \citenamefont {Bashmakova}, \citenamefont {Tagantsev},\ and\ \citenamefont {Golubeva}}]{korolev2024generation}%
  \BibitemOpen
  \bibfield  {author} {\bibinfo {author} {\bibfnamefont {S.}~\bibnamefont {Korolev}}, \bibinfo {author} {\bibfnamefont {E.}~\bibnamefont {Bashmakova}}, \bibinfo {author} {\bibfnamefont {A.}~\bibnamefont {Tagantsev}},\ and\ \bibinfo {author} {\bibfnamefont {T.~Y.}\ \bibnamefont {Golubeva}},\ }\href {https://doi.org/10.1103/PhysRevA.109.052428} {\bibfield  {journal} {\bibinfo  {journal} {Physical Review A}\ }\textbf {\bibinfo {volume} {109}},\ \bibinfo {pages} {052428} (\bibinfo {year} {2024})}\BibitemShut {NoStop}%
\bibitem [{\citenamefont {Kienzler}\ \emph {et~al.}(2017)\citenamefont {Kienzler}, \citenamefont {Lo}, \citenamefont {Negnevitsky}, \citenamefont {Fl{\"u}hmann}, \citenamefont {Marinelli},\ and\ \citenamefont {Home}}]{kienzler2017quantum}%
  \BibitemOpen
  \bibfield  {author} {\bibinfo {author} {\bibfnamefont {D.}~\bibnamefont {Kienzler}}, \bibinfo {author} {\bibfnamefont {H.-Y.}\ \bibnamefont {Lo}}, \bibinfo {author} {\bibfnamefont {V.}~\bibnamefont {Negnevitsky}}, \bibinfo {author} {\bibfnamefont {C.}~\bibnamefont {Fl{\"u}hmann}}, \bibinfo {author} {\bibfnamefont {M.}~\bibnamefont {Marinelli}},\ and\ \bibinfo {author} {\bibfnamefont {J.}~\bibnamefont {Home}},\ }\href {https://doi.org/10.1103/PhysRevLett.119.033602} {\bibfield  {journal} {\bibinfo  {journal} {Physical review letters}\ }\textbf {\bibinfo {volume} {119}},\ \bibinfo {pages} {033602} (\bibinfo {year} {2017})}\BibitemShut {NoStop}%
\bibitem [{\citenamefont {Bashmakova}\ \emph {et~al.}(2025)\citenamefont {Bashmakova}, \citenamefont {Korolev},\ and\ \citenamefont {Golubeva}}]{bashmakova2025bosonic}%
  \BibitemOpen
  \bibfield  {author} {\bibinfo {author} {\bibfnamefont {E.}~\bibnamefont {Bashmakova}}, \bibinfo {author} {\bibfnamefont {S.}~\bibnamefont {Korolev}},\ and\ \bibinfo {author} {\bibfnamefont {T.~Y.}\ \bibnamefont {Golubeva}},\ }\href {https://doi.org/10.1103/97yt-nzg2} {\bibfield  {journal} {\bibinfo  {journal} {Physical Review A}\ }\textbf {\bibinfo {volume} {112}},\ \bibinfo {pages} {032434} (\bibinfo {year} {2025})}\BibitemShut {NoStop}%
\bibitem [{\citenamefont {Cai}\ \emph {et~al.}(2025)\citenamefont {Cai}, \citenamefont {Deng}, \citenamefont {Zhang}, \citenamefont {Ni}, \citenamefont {Mai}, \citenamefont {Huang}, \citenamefont {Zheng}, \citenamefont {Hu}, \citenamefont {Liu}, \citenamefont {Xu} \emph {et~al.}}]{cai2025quantum}%
  \BibitemOpen
  \bibfield  {author} {\bibinfo {author} {\bibfnamefont {Y.}~\bibnamefont {Cai}}, \bibinfo {author} {\bibfnamefont {X.}~\bibnamefont {Deng}}, \bibinfo {author} {\bibfnamefont {L.}~\bibnamefont {Zhang}}, \bibinfo {author} {\bibfnamefont {Z.}~\bibnamefont {Ni}}, \bibinfo {author} {\bibfnamefont {J.}~\bibnamefont {Mai}}, \bibinfo {author} {\bibfnamefont {P.}~\bibnamefont {Huang}}, \bibinfo {author} {\bibfnamefont {P.}~\bibnamefont {Zheng}}, \bibinfo {author} {\bibfnamefont {L.}~\bibnamefont {Hu}}, \bibinfo {author} {\bibfnamefont {S.}~\bibnamefont {Liu}}, \bibinfo {author} {\bibfnamefont {Y.}~\bibnamefont {Xu}}, \emph {et~al.},\ }\bibfield  {journal} {\bibinfo  {journal} {arXiv preprint arXiv:2503.08197}\ }\href {https://doi.org/10.48550/arXiv.2503.08197} {10.48550/arXiv.2503.08197} (\bibinfo {year} {2025})\BibitemShut {NoStop}%
\bibitem [{\citenamefont {Zeng}\ \emph {et~al.}(2025)\citenamefont {Zeng}, \citenamefont {Quijandr{\'\i}a}, \citenamefont {Gneiting},\ and\ \citenamefont {Nori}}]{zeng2025quantum}%
  \BibitemOpen
  \bibfield  {author} {\bibinfo {author} {\bibfnamefont {Y.}~\bibnamefont {Zeng}}, \bibinfo {author} {\bibfnamefont {F.}~\bibnamefont {Quijandr{\'\i}a}}, \bibinfo {author} {\bibfnamefont {C.}~\bibnamefont {Gneiting}},\ and\ \bibinfo {author} {\bibfnamefont {F.}~\bibnamefont {Nori}},\ }\bibfield  {journal} {\bibinfo  {journal} {arXiv preprint arXiv:2510.04209}\ }\href {https://doi.org/10.48550/arXiv.2510.04209} {10.48550/arXiv.2510.04209} (\bibinfo {year} {2025})\BibitemShut {NoStop}%
\bibitem [{\citenamefont {Winnel}\ \emph {et~al.}(2024)\citenamefont {Winnel}, \citenamefont {Guanzon}, \citenamefont {Singh},\ and\ \citenamefont {Ralph}}]{winnel2024deterministic}%
  \BibitemOpen
  \bibfield  {author} {\bibinfo {author} {\bibfnamefont {M.~S.}\ \bibnamefont {Winnel}}, \bibinfo {author} {\bibfnamefont {J.~J.}\ \bibnamefont {Guanzon}}, \bibinfo {author} {\bibfnamefont {D.}~\bibnamefont {Singh}},\ and\ \bibinfo {author} {\bibfnamefont {T.~C.}\ \bibnamefont {Ralph}},\ }\href {https://doi.org/10.1103/PhysRevLett.132.230602} {\bibfield  {journal} {\bibinfo  {journal} {Physical Review Letters}\ }\textbf {\bibinfo {volume} {132}},\ \bibinfo {pages} {230602} (\bibinfo {year} {2024})}\BibitemShut {NoStop}%
\bibitem [{\citenamefont {Peralta}\ \emph {et~al.}(2013)\citenamefont {Peralta}, \citenamefont {Soong}, \citenamefont {England}, \citenamefont {Colby}, \citenamefont {Wu}, \citenamefont {Montazeri}, \citenamefont {McGuinness}, \citenamefont {McNeur}, \citenamefont {Leedle}, \citenamefont {Walz} \emph {et~al.}}]{peralta2013demonstration}%
  \BibitemOpen
  \bibfield  {author} {\bibinfo {author} {\bibfnamefont {E.}~\bibnamefont {Peralta}}, \bibinfo {author} {\bibfnamefont {K.}~\bibnamefont {Soong}}, \bibinfo {author} {\bibfnamefont {R.}~\bibnamefont {England}}, \bibinfo {author} {\bibfnamefont {E.}~\bibnamefont {Colby}}, \bibinfo {author} {\bibfnamefont {Z.}~\bibnamefont {Wu}}, \bibinfo {author} {\bibfnamefont {B.}~\bibnamefont {Montazeri}}, \bibinfo {author} {\bibfnamefont {C.}~\bibnamefont {McGuinness}}, \bibinfo {author} {\bibfnamefont {J.}~\bibnamefont {McNeur}}, \bibinfo {author} {\bibfnamefont {K.}~\bibnamefont {Leedle}}, \bibinfo {author} {\bibfnamefont {D.}~\bibnamefont {Walz}}, \emph {et~al.},\ }\href {https://doi.org/10.1038/nature12664} {\bibfield  {journal} {\bibinfo  {journal} {Nature}\ }\textbf {\bibinfo {volume} {503}},\ \bibinfo {pages} {91} (\bibinfo {year} {2013})}\BibitemShut {NoStop}%
\bibitem [{\citenamefont {Sapra}\ \emph {et~al.}(2020)\citenamefont {Sapra}, \citenamefont {Yang}, \citenamefont {Vercruysse}, \citenamefont {Leedle}, \citenamefont {Black}, \citenamefont {England}, \citenamefont {Su}, \citenamefont {Trivedi}, \citenamefont {Miao}, \citenamefont {Solgaard} \emph {et~al.}}]{sapra2020chip}%
  \BibitemOpen
  \bibfield  {author} {\bibinfo {author} {\bibfnamefont {N.~V.}\ \bibnamefont {Sapra}}, \bibinfo {author} {\bibfnamefont {K.~Y.}\ \bibnamefont {Yang}}, \bibinfo {author} {\bibfnamefont {D.}~\bibnamefont {Vercruysse}}, \bibinfo {author} {\bibfnamefont {K.~J.}\ \bibnamefont {Leedle}}, \bibinfo {author} {\bibfnamefont {D.~S.}\ \bibnamefont {Black}}, \bibinfo {author} {\bibfnamefont {R.~J.}\ \bibnamefont {England}}, \bibinfo {author} {\bibfnamefont {L.}~\bibnamefont {Su}}, \bibinfo {author} {\bibfnamefont {R.}~\bibnamefont {Trivedi}}, \bibinfo {author} {\bibfnamefont {Y.}~\bibnamefont {Miao}}, \bibinfo {author} {\bibfnamefont {O.}~\bibnamefont {Solgaard}}, \emph {et~al.},\ }\href {https://doi.org/10.1126/science.aay5734} {\bibfield  {journal} {\bibinfo  {journal} {Science}\ }\textbf {\bibinfo {volume} {367}},\ \bibinfo {pages} {79} (\bibinfo {year} {2020})}\BibitemShut {NoStop}%
\bibitem [{\citenamefont {Shiloh}\ \emph {et~al.}(2021)\citenamefont {Shiloh}, \citenamefont {Illmer}, \citenamefont {Chlouba}, \citenamefont {Yousefi}, \citenamefont {Sch{\"o}nenberger}, \citenamefont {Niedermayer}, \citenamefont {Mittelbach},\ and\ \citenamefont {Hommelhoff}}]{shiloh2021electron}%
  \BibitemOpen
  \bibfield  {author} {\bibinfo {author} {\bibfnamefont {R.}~\bibnamefont {Shiloh}}, \bibinfo {author} {\bibfnamefont {J.}~\bibnamefont {Illmer}}, \bibinfo {author} {\bibfnamefont {T.}~\bibnamefont {Chlouba}}, \bibinfo {author} {\bibfnamefont {P.}~\bibnamefont {Yousefi}}, \bibinfo {author} {\bibfnamefont {N.}~\bibnamefont {Sch{\"o}nenberger}}, \bibinfo {author} {\bibfnamefont {U.}~\bibnamefont {Niedermayer}}, \bibinfo {author} {\bibfnamefont {A.}~\bibnamefont {Mittelbach}},\ and\ \bibinfo {author} {\bibfnamefont {P.}~\bibnamefont {Hommelhoff}},\ }\href {https://doi.org/10.1038/s41586-021-03812-9} {\bibfield  {journal} {\bibinfo  {journal} {Nature}\ }\textbf {\bibinfo {volume} {597}},\ \bibinfo {pages} {498} (\bibinfo {year} {2021})}\BibitemShut {NoStop}%
\bibitem [{\citenamefont {Adiv}\ \emph {et~al.}(2021)\citenamefont {Adiv}, \citenamefont {Wang}, \citenamefont {Dahan}, \citenamefont {Broaddus}, \citenamefont {Miao}, \citenamefont {Black}, \citenamefont {Leedle}, \citenamefont {Byer}, \citenamefont {Solgaard}, \citenamefont {England} \emph {et~al.}}]{adiv2021quantum}%
  \BibitemOpen
  \bibfield  {author} {\bibinfo {author} {\bibfnamefont {Y.}~\bibnamefont {Adiv}}, \bibinfo {author} {\bibfnamefont {K.}~\bibnamefont {Wang}}, \bibinfo {author} {\bibfnamefont {R.}~\bibnamefont {Dahan}}, \bibinfo {author} {\bibfnamefont {P.}~\bibnamefont {Broaddus}}, \bibinfo {author} {\bibfnamefont {Y.}~\bibnamefont {Miao}}, \bibinfo {author} {\bibfnamefont {D.}~\bibnamefont {Black}}, \bibinfo {author} {\bibfnamefont {K.}~\bibnamefont {Leedle}}, \bibinfo {author} {\bibfnamefont {R.~L.}\ \bibnamefont {Byer}}, \bibinfo {author} {\bibfnamefont {O.}~\bibnamefont {Solgaard}}, \bibinfo {author} {\bibfnamefont {R.~J.}\ \bibnamefont {England}}, \emph {et~al.},\ }\href {https://doi.org/10.1103/PhysRevX.11.041042} {\bibfield  {journal} {\bibinfo  {journal} {Physical Review X}\ }\textbf {\bibinfo {volume} {11}},\ \bibinfo {pages} {041042} (\bibinfo {year} {2021})}\BibitemShut {NoStop}%
\bibitem [{\citenamefont {Chlouba}\ \emph {et~al.}(2023)\citenamefont {Chlouba}, \citenamefont {Shiloh}, \citenamefont {Kraus}, \citenamefont {Br{\"u}ckner}, \citenamefont {Litzel},\ and\ \citenamefont {Hommelhoff}}]{chlouba2023coherent}%
  \BibitemOpen
  \bibfield  {author} {\bibinfo {author} {\bibfnamefont {T.}~\bibnamefont {Chlouba}}, \bibinfo {author} {\bibfnamefont {R.}~\bibnamefont {Shiloh}}, \bibinfo {author} {\bibfnamefont {S.}~\bibnamefont {Kraus}}, \bibinfo {author} {\bibfnamefont {L.}~\bibnamefont {Br{\"u}ckner}}, \bibinfo {author} {\bibfnamefont {J.}~\bibnamefont {Litzel}},\ and\ \bibinfo {author} {\bibfnamefont {P.}~\bibnamefont {Hommelhoff}},\ }\href {https://doi.org/10.1038/s41586-023-06602-7} {\bibfield  {journal} {\bibinfo  {journal} {Nature}\ }\textbf {\bibinfo {volume} {622}},\ \bibinfo {pages} {476} (\bibinfo {year} {2023})}\BibitemShut {NoStop}%
\end{thebibliography}
\end{document}